\documentclass[10pt,conference,letterpaper]{IEEEtran}
\IEEEoverridecommandlockouts

\usepackage{epsfig}
\usepackage{tikz}
\usepackage{amsmath,tabu}
\usepackage{graphicx}
\usepackage{adjustbox}
\usepackage{amssymb,amsmath,enumitem}
\usepackage{fancyvrb}
\usepackage{color,soul}
\usepackage{array,colortbl,xcolor}
\usepackage{pgfplots}
\usepackage{pgfplotstable}
\usepackage{url}
\usepackage{subcaption}
\usepackage{fancyvrb}

\usepackage{listings}
\def\BibTeX{{\rm B\kern-.05em{\sc i\kern-.025em b}\kern-.08em
    T\kern-.1667em\lower.7ex\hbox{E}\kern-.125emX}}

\newbox\verbbox
\usetikzlibrary{patterns}
\usetikzlibrary{decorations.pathmorphing}
\usetikzlibrary{positioning,calc, fit}
\usetikzlibrary{shapes,arrows,shadows, shapes.multipart}

\definecolor{myorange}{RGB}{250,192,116}
\definecolor{rred}{HTML}{C0504D}
\definecolor{lgreen}{RGB}{204, 204, 154}
\definecolor{ppurple}{HTML}{9F4C7C}
\definecolor{ggreen}{HTML}{9BBB59}
\definecolor{bblue}{HTML}{4F81BD}
\definecolor{palepurple}{HTML}{F1DAFF}
\definecolor{palemint}{HTML}{CCFFDD}
\definecolor{palegreen}{HTML}{CCEEDD}
\definecolor{palered}{HTML}{FFCCCC}
\definecolor{pastelteal}{HTML}{A6DDCA}
\definecolor{pastelpurple}{HTML}{C7B7D9}

\definecolor{ref}{rgb}{0.65,0.65,0.65} 
\definecolor{lhmm}{rgb}{0.9,0.6,0.5}
\definecolor{ghmm}{rgb}{0.7,0.9,0.35}
\definecolor{lsvm}{rgb}{0.9,0.8,0.25}
\definecolor{gsvm}{rgb}{0.4,0.8,0.9}

\pgfplotsset{compat=newest}

\pgfkeys{
    /tikz/node distance/.append code={
        \pgfkeyssetvalue{/tikz/node distance value}{#1}
    }
}

\newcommand\widernode[5][myorange]{
\node[
        #1,
        inner sep=0pt,
        shift=($(#2.south)-(#2.north)$),
        yshift=-\pgfkeysvalueof{/tikz/node distance value},
        fit={(#2) (#3)},
        label=center:{\scriptsize#4}] (#5) {};
}

\makeatletter

\def\pgf@sh@bg@tape{
    \tapedimensions%
    \pgf@xc\halfwidth%
    \pgf@yc\halfheight%
    \pgf@xc\bendxradius%
    \pgf@yc\bendyradius%
    {%
      \pgftransformshift{\centerpoint}%
      \pgfpathmoveto{\pgfqpoint{-\halfwidth}{0pt}}%
      \pgfpathlineto{\pgfqpoint{-\halfwidth}{\halfheight}}%
      \ifx\topbendstyle\pgf@lib@sh@inandouttext%
        \pgfpathlineto{\pgf@x-\halfwidth\pgf@y\halfheight\advance\pgf@y\halfbendheight}%
        \pgfpatharc{225}{315}{\bendxradius and \bendyradius}%
        \pgfpatharc{135}{45}{\bendxradius and \bendyradius}%
      \else%
        \ifx\topbendstyle\pgf@lib@sh@outandintext%
          \pgfpathlineto{\pgf@x-\halfwidth\pgf@y\halfheight\advance\pgf@y\halfbendheight}%
          \pgfpatharc{135}{45}{\bendxradius and \bendyradius}%
          \pgfpatharc{225}{315}{\bendxradius and \bendyradius}%
        \else%
          \pgfpathlineto{\pgfqpoint{\halfwidth}{\halfheight}}%
        \fi%
      \fi%
      \ifx\bottombendstyle\pgf@lib@sh@inandouttext%
        \pgfpathlineto{\pgf@x\halfwidth\pgf@y-\halfheight\advance\pgf@y-\halfbendheight}%
        \pgfpatharc{45}{135}{\bendxradius and \bendyradius}%
        \pgfpatharc{315}{225}{\bendxradius and \bendyradius}%
      \else%
        \ifx\bottombendstyle\pgf@lib@sh@outandintext%
          \pgfpathlineto{\pgf@x\halfwidth\pgf@y-\halfheight\advance\pgf@y-\halfbendheight}%
          \pgfpatharc{315}{225}{\bendxradius and \bendyradius}%
          \pgfpatharc{45}{135}{\bendxradius and \bendyradius}%
        \else%
          \pgfpathlineto{\pgfqpoint{\halfwidth}{-\halfheight}}%
          \pgfpathlineto{\pgfqpoint{-\halfwidth}{-\halfheight}}%
        \fi%
      \fi%
      \pgfpathclose%
    }%
}

\tikzset{ 
    docushape/.style={
        shape=tape,
        minimum width=5cm,
        inner ysep=3pt,
        draw,
        ultra thick,
        align=center,
        fill=white,
        font=\scriptsize\ttfamily, 
        tape bend top=none,
    },
    arrowstyle/.style={
        single arrow, 
        fill=blue!50, 
        anchor=base, 
        align=center,
        ultra thick,
        text width=2.8cm
    },
    mymodule/.style={
        rounded corners=1mm,
        draw=black,
        ultra thick,
        font=\scriptsize\ttfamily, 
        minimum width=2cm,
        minimum height=0.5cm,
        inner ysep=1pt
    },
    myshade/.style={
       rounded corners=1mm,
       ultra thick,
       draw=gray,
       font=\scriptsize\ttfamily, 
       inner ysep=4pt,
       inner xsep=4pt
    }
}

\def\pname{FRAMER}

\begin{document}

%
%
\setlength{\textfloatsep}{1pt plus 0.0pt minus 0.0pt}
\setlength{\dbltextfloatsep}{0.5pt}

\date{}


\title{
    \pname{}: A Software-based Capability Model
}
\author{

    \IEEEauthorblockN{1\textsuperscript{st} Myoung Jin Nam}
    \IEEEauthorblockA{
        \textit{Korea University}\\
            Seoul, S.Korea \\
            mjnam@formal.korea.ac.kr}
    \and

    \IEEEauthorblockN{2\textsuperscript{nd} Periklis Akritidis}
    \IEEEauthorblockA{
    \textit{Niometrics}\\
           Singapore, Singapore \\
           akritid@niometrics.com}
    
    \and

    \IEEEauthorblockN{3\textsuperscript{rd} David J Greaves}
    \IEEEauthorblockA{
    \textit{University of Cambridge}\\
            Cambridge, UK\\
            djg11@cl.cam.ac.uk}
    
   
}


\maketitle


\subsection*{Abstract}
Fine-grained memory protection for C and C++ programs 
must track individual objects (or pointers), and 
store bounds information per object (or pointer).
Its cost is dominated by metadata updates and lookups,
making efficient metadata management the key for minimizing
performance impact.
Existing approaches reduce metadata management 
overheads by sacrificing precision, 
breaking binary compatibility by changing object memory layout, 
or wasting space with excessive alignment 
or large shadow memory spaces.

We propose \pname{}, a software capability model
with object granularity. Its efficient 
per-object metadata management mechanism enables 
direct access to metadata by calculating their location 
from a \textit{tagged pointer} to the object
and a compact supplementary table,
and has potential applications 
in memory safety, type safety, thread safety and garbage 
collection.
\pname{} improves over previous solutions by 
simultaneously
(1) streamlining expensive metadata lookups, 
(2) offering flexibility in metadata placement and size,
(3) saving space by removing superfluous alignment and padding, and
(4) avoiding internal object memory layout changes. 
We evaluate \pname{} with a use case on memory protection.


\section{Introduction}

Despite advances in software defenses, exploitation of
systems code written in unsafe languages such as C and C++
is still possible.
Security exploits use memory safety vulnerabilities
to corrupt or leak sensitive data,
and hijack a vulnerable program's logic.
In response, several defenses have been proposed 
for making software exploitation hard.

Current defenses fall in two basic categories: 
those that let memory corruption happen, but harden the 
program to prevent exploitation, and those that try to
detect and block memory corruption in the first place. 
For instance, Control-flow Integrity (CFI)~\cite{cfi,tvip,safedispatch,mocfi,vfguard,vtv,patharmor,typearmor,vtrust,ccfir,bincfi}
contains all control flows
in a statically computed 
Control-flow Graph (CFG), while Address Space Layout Randomization (ASLR)
hides the available
CFG when the process executes. Both approaches can
offer only probabilistic security~\cite{out-of-control,jitrop},
since memory corruption 
is still possible, albeit exploitation is much harder. 



A general approach to detect and block memory corruption
is through tracking the bounds of object allocations~\cite{Akritidis:2009:BBC:1855768.1855772,
Austin:1994:EDP:178243.178446,
Duck:2018:ETM:3192366.3192388,
typesan,
intel_mpx,
hextype,
Jones97backwards-compatiblebounds,
Kroes:2018:DPB:3190508.3190553,
softbound,
cets,
Necula:2005:CTR:1065887.1065892,
address-sanitizer,
Younan:2010:PEP:1755688.1755707}.
The program is instrumented
accordingly to use bounds information for 
blocking unintended accesses to objects.
These systems can offer 
deterministic guarantees, since now memory corruption is
prevented in the first place, 
however tracking all objects (or pointers)
incurs heavy performance overheads. 

Some existing techniques trade off \textit{compatibility} for 
high locality of reference, however, it is desirable to minimise 
the disruption owing to tacit assumptions by programmers and 
compatibility with existing code or libraries that cannot be 
recompiled.
In particular, so-called \textit{fat pointers}
\cite{Necula:2005:CTR:1065887.1065892}
impose incompatibility issues with external 
modules, especially pre-compiled libraries in software-based   
solutions.

With these limitations in mind,
\textit{object-capability models} 
\cite{Dennis65programmingsemantics,Kwon13low-fatpointers:,
7163016,2665740}
using hardware-supported
tags become very attractive, because they can manage 
compatibility and control runtime costs.
But they are not supported in today's mainstream processor architectures,
and, more importantly,
cannot entirely avoid undesirable overheads such as metadata management
related memory accesses just by virtue of being hardware-based.


In this paper, we present \pname{}, a software-based capability 
model using \textit{tagged pointers} for fast metadata access.
\pname{} provides efficient and flexible per-object metadata management 
that enables direct access to metadata by calculating 
their location using the (currently) unused top 16 bits of a 64-bit
pointer to the object and a supplementary table.
The key considerations behind \pname{} are as follows.


Firstly, 
\pname{} enables the memory manager freedom to place
metadata in the associated header near the object 
to maximise spatial locality,
which has positive effects at all levels of the memory hierarchy.
Headers can vary in size, unlike approaches that store 
the header at a system-wide fixed offset from the object,
which may be useful in some applications.
Headers can also be shared over object instances (although we do
not develop that aspect in this paper).
Our evaluation shows excellent D-cache performance where
the performance impact of software checking is, to a fair extent,
mitigated by improved instructions per cycle (IPC).

Secondly, the address of the header holding metadata is
derived from tagged pointers regardless of objects' alignment. 
We use a novel technique to encode the \textit{relative location} of the header 
in unused bits at the top of a pointer.
This streamlines metadata lookup,
which has been the performance bottleneck of deterministic 
memory safety approaches. Moreover, the encoding is such,
that despite being relative to the address in the pointer,
the tag does not require updating
when the address in the pointer changes.
A supplementary table is used
only for cases where the location information cannot 
be directly addressed with the additional 16-bits 
in the pointer. The address of the corresponding entry 
in the table is also calculated from our tagged pointer.   
This table is small compared to typical \textit{shadow memory} 
implementations.

Thirdly, we avoid wasting memory from excessive 
padding and superfluous alignment, 
by encoding and using relative location for metadata access. 
Whereas existing approaches using shadow space~\cite{Akritidis:2009:BBC:1855768.1855772,
DBLP:conf/eurosec/HallerKGB16,
Kwon13low-fatpointers:,softbound,address-sanitizer}
re-align or group objects to avoid conflicts in entries, 
\pname{} provides great flexibility in alignment, that 
completely removes constraining the objects or memory. 
The average of space overheads of our approach is 
20\% for full checking despite the generous size
of metadata in our current implementation.

Fourthly, our approach facilitates \textit{compatibility}.
Our tag is encoded in otherwise unused bits 
at the top of a pointer, but the pointer size 
is unchanged and contiguity can be ensured. 



The contributions of this paper are the following:
\begin{itemize}

\item We present an encoding technique
for relative offsets that is interesting in its own right.
It is both compact and also avoids imposing
object alignment or size constraints.
Moreover, it is favourable for hardware implementation
and may find uses across different application domains.

\item We design, implement and evaluate \pname{}, 
a generic framework for fast and practical 
object-metadata management with potential applications 
in memory safety, type safety and garbage collection. 

\item We present a case study of applying \pname{} 
to the problem of spatial memory safety,
using our framework to allow inexpensive validation
of pointer dereferences.
We further discuss some kinds of
violations of temporal safety that \pname{} prevents,
and a potential application for type confusion prevention.

\end{itemize}

\section{Background and Related Work}


In this section we discuss techniques to detect memory errors.
Static analysis detects errors at compile time and does not
introduce run-time overhead, but must inevitably be 
over-conservative, giving false alarms.
In this paper, we focus only on run-time verification, 
but with compile-time assistance.

\subsection{Metadata Association}

Several approaches have been proposed for tracking memory 
and detecting memory-related errors. We review here systems 
that either track objects or pointers. 

\subsubsection{Object-based Tracking}

An \textit{object-based} approach stores bounds information 
per object. 
By not changing the memory layout of objects, it offers compatibility 
with current source and pre-compiled legacy libraries. 
In addition to checking pointer dereferences,
these approaches may check pointer arithmetic to avoid
losing track of the \textit{intended referent}~\cite{Jones97backwards-compatiblebounds},
or otherwise risk false negatives for some spatial violations
that stride over object boundaries.
Moreover, object-based approaches checking pointer arithmetic
should take special 
care to avoid false positives for valid out-of-bound pointers
that are never dereferenced.
Object-based approaches usually omit tracking of sub-objects 
such as an array member of a structure. 

Due to the arbitrary size of objects, these approaches 
require some form of range-based lookup of objects using an in-bound address,
which can be more expensive than a simple lookup.
An early approach used a splay tree to reduce the 
overhead \cite{Jones97backwards-compatiblebounds},
but more efficient systems simplify the lookup by using table access
to shadow memory regions.
We will discuss the details in Section~\ref{sec:metadata_storage}.

\subsubsection{Pointer-based tracking}

\textit{Pointer-based} approaches associate bounds 
information with pointers. Per-pointer metadata
holds the valid range that a pointer is allowed to point to.  
Unlike object-based approaches,
this enables them to detect internal overflows easier,
such as an array out-of-bounds inside a structure, 
so pointer-based approaches
can guarantee near \textit{complete memory safety}. 
One drawback is the additional runtime overhead from metadata
copy and update at pointer assignment, while object-based 
approaches update metadata only at memory allocation/release.
In addition, the number of pointers can be larger than 
that of allocated objects, so pointer-intensive programs
may suffer from heavier runtime overheads.   

\subsection{Metadata Storage}\label{sec:metadata_storage}

Memory safety enforcement techniques fall into two 
categories depending on whether metadata is \textit{disjoint} 
or \textit{embedded} in each object or pointer.

\subsubsection{Embedded Metadata}\label{sec:embedded_metadata}

\paragraph{Fat Pointers}

Fat pointers~\cite{Austin:1994:EDP:178243.178446, Jim_cyclone:a, 
Necula:2005:CTR:1065887.1065892} embed metadata in each 
pointer as shown in Fig.~\ref{fig:fatpointer}. 
They provide speed with the highest locality of 
references by avoiding extra memory access for metadata 
update/retrieval, however, they break binary compatibility due to 
expansion of the pointer representation. In addition, fat 
pointers are vulnerable to metadata corruption by store
operations after unsafe typecast operations on pointers. 
Hence, it is essential to check typecasts to guarantee 
near-complete memory safety.  

\tikzstyle{arrow}=[draw, -latex] 

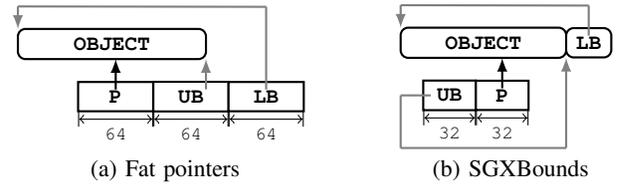
\begin{figure}[t!]
  \centering


  \begin{subfigure}[t]{0.5\columnwidth}
  \centering
  \begin{tikzpicture}[
    node distance=0pt,
    box/.style={
        node distance=0pt,outer sep=0pt,
        draw=black,
        thick,
        fill=white, 
        minimum height=0.3cm,
        font=\fontsize{8}{8}\ttfamily, 
        anchor=west
    },
  ]
  
  \node[box=1.9cm, minimum width=2.5cm, rounded corners=1mm](fatptr_obj) 
        at (-1.5, 1.3)   
        {\textbf{OBJECT}};
  
  \node[box=0.4cm, minimum width=1.0cm](fatptr_ptr) 
        at (-0.7, 0.6) 
        {\textbf{P}};
  
  \draw[|<->|] ([yshift= -0.3cm] fatptr_ptr.west) 
                -- node[below] {\scriptsize\ttfamily 64} 
                ([yshift= -0.3cm]fatptr_ptr.east);
  
  \node[box=0.4cm, minimum width=1.0cm, right=of fatptr_ptr] (fatptr_ub) 
        {\textbf{UB}};  
         
  \draw[|<->|] ([yshift= -0.3cm] fatptr_ub.west) 
                -- node[below] {\scriptsize\ttfamily 64} 
                ([yshift= -0.3cm]fatptr_ub.east);
  
  \node[box=0.4cm, minimum width=1.0cm, right=of fatptr_ub] (fatptr_lb) 
        {\textbf{LB}};   
  
  \draw[|<->|] ([yshift= -0.3cm] fatptr_lb.west) 
                -- node[below] {\scriptsize\ttfamily 64} 
                ([yshift= -0.3cm]fatptr_lb.east);
  
  \path[arrow, thick] ([yshift= -0.1cm] fatptr_ptr.north) 
                -- 
               ([xshift= 0.05cm]fatptr_obj.south);
  
  \path[arrow, thick, gray] ([xshift= 0.2cm, yshift= -0.1cm] fatptr_ub.north) 
                -- 
               ([yshift= -0.2cm]fatptr_obj.east);

  \draw[-, thick, gray] ([yshift= -0.1cm] fatptr_lb.north) 
                -- 
           ([yshift= 1.0cm]fatptr_lb.north);
  
  \draw[-, thick, gray] ([yshift= 1.0cm]fatptr_lb.north) 
                -- 
           ([yshift= 0.5cm]fatptr_obj.west);
  \path[arrow, thick, gray] ([yshift= 0.5cm]fatptr_obj.west) 
                -- 
               ([yshift= 0.2cm]fatptr_obj.west);
  \end{tikzpicture}
  \vspace*{-1mm}
  \caption{\small{Fat pointers}}
  \label{fig:fatpointer}
  \end{subfigure}%
  ~
  \begin{subfigure}[t]{0.5\columnwidth}
  \centering
  \begin{tikzpicture}[
    node distance=0pt,
    box/.style={
        node distance=0pt,outer sep=0pt,
        draw=black,
        thick,
        fill=white, 
        minimum height=0.3cm,
        font=\fontsize{8}{8}\ttfamily, 
        anchor=west,
    },
  ]

  \node[box=1.9cm, minimum width=2.2cm, 
        rounded corners=1mm](SGX_obj) 
      at (-1.5, -0.8) 
       {\textbf{OBJECT}};
  
  \node[box=0.40cm, minimum width=0.6cm, 
        right=of SGX_obj, rounded corners=1mm](SGX_lb)   
        {\textbf{LB}};

  \node[box=0.4cm, minimum width=0.7cm](SGX_ub) 
      at (-1.2, -1.5) 
      {\textbf{UB}};
  
  \draw[|<->|] ([yshift= -0.3cm] SGX_ub.west) 
                -- node[below] {\scriptsize\ttfamily 32} 
                ([yshift= -0.3cm]SGX_ub.east);
  
  \node[box=0.4cm, minimum width=0.7cm, right=of SGX_ub](SGX_ptr) 
          {\textbf{P}};   
  
  \draw[|<->|] ([yshift= -0.3cm] SGX_ptr.west) 
                -- node[below] {\scriptsize\ttfamily 32} 
                ([yshift= -0.3cm] SGX_ptr.east);
  
  \path[arrow, thick] ([yshift= -0.1cm] SGX_ptr.north) 
                 -- 
               ([xshift =0.25cm]SGX_obj.south);
   
  \draw[-, thick, gray] ([xshift= 0.1cm] SGX_ub.west) 
                 -- 
           ([xshift =-0.3cm] SGX_ub.west);
  \draw[-, thick, gray] ([xshift= -0.3cm] SGX_ub.west) 
                 -- 
           ([xshift =-0.3cm, yshift=-0.7cm] SGX_ub.west);
  \draw[-, thick, gray] ([xshift= -0.3cm, yshift=-0.7cm] SGX_ub.west) 
                 -- 
           ([yshift =-1.4cm] SGX_obj.east);
  \path[arrow, thick, gray]([yshift =-1.4cm] SGX_obj.east) 
                 -- 
              ([yshift =-0.2cm] SGX_obj.east);
  
  \draw[-, thick, gray] ([yshift= -0.1cm] SGX_lb.north) 
                 -- 
           ([yshift =0.3cm] SGX_lb.north);
   
  \draw[-, thick, gray] ([yshift= 0.3cm] SGX_lb.north) 
                 -- 
           ([yshift =0.5cm] SGX_obj.west);
  \path[arrow, thick, gray] ([yshift= 0.5cm] SGX_obj.west) 
                 -- 
            ([yshift =0.2cm] SGX_obj.west);
    
    \end{tikzpicture}
  \vspace*{-1mm}
  \caption{\small{SGXBounds}}
  \label{fig:sgxbounds}
  \end{subfigure}%
\vspace*{1mm}
\caption{\small{Embedded Metadata: 
P, UB, and LB represent a pointer itself, 
upper bound, and lower bound, respectively.}}
\label{comparison1}
\end{figure}

\paragraph{Tagged Pointers}
To avoid fat pointers, several techniques
using embedded metadata employ tagged pointers instead.
SGXBounds~\cite{Kuvaiskii:2017:SMS:3064176.3064192} utilizes 
tagged pointers like \pname{} and makes objects carry
their metadata in a \textit{footer} as shown 
in Fig.~\ref{fig:sgxbounds}. In SGXBounds, 
a 64-bit pointer's lower 32 bits hold the address, and the higher 
32 bits hold the referent object's \textit{upper bound}, 
i.e.~the location of the metadata footer. The footer contains 
a \textit{lower bound} (base), and may hold other metadata.
This approach works when there are enough spare bits in pointers,
which is the case with SGX enclaves, where only 36 bits of virtual address 
space are currently supported.
Storing the absolute address of bounds frees SGXBounds 
from false negatives/positives that challenge many object-tracking 
approaches.
The use of a footer provides fairly high locality of
references, and is less vulnerable to metadata corruption by
unsafe typecast than fat pointers. 

Other techniques, such as Baggy Bounds Checking~\cite{Akritidis:2009:BBC:1855768.1855772} and
Low-fat Pointers~\cite{Kwon13low-fatpointers:} use different compromises
to support larger address spaces without changing the pointer size.

\subsubsection{Disjoint Metadata}\label{sec:disjoint_metadata}

Disjoint metadata achieves memory layout compatibility
by storing metadata in a separate memory region.
A high-level data structure, such as a hash table, simplifies 
implementation and manipulation of metadata, however, runtime   
overheads can be lower when using a shadow space that allows 
direct array access to metadata~\cite{Akritidis:2009:BBC:1855768.1855772,
Cheng:2006:TEF:1157733.1157903,
Devietti:2008:HAS:1353536.1346295,
DBLP:conf/eurosec/HallerKGB16,
softbound,
Necula:2005:CTR:1065887.1065892,4041842,
Xu:2004:EBT:1029894.1029913}. 
Early techniques using shadow spaces create a mirror copy of application space,
i.e.~byte-to-byte mapping, but more recent techniques 
reduced the size of shadow space with compact encoding, 
re-arranging objects and so on.



SoftBound~\cite{softbound} 
is a pointer-based approach that
uses either a hash table and shadow space, while ensuring
compatibility and protecting sub-objects, and shows that 
the use of shadow space reduces runtime overhead, on average, 
by 2/3 compared with using table lookup~\cite{DBLP:conf/snapl/NagarakatteMZ15}.




Baggy bounds checking (BBC)~\cite{Akritidis:2009:BBC:1855768.1855772}
is an object-based approach that includes an 
implementation using shadow memory that maps
fixed-sized memory blocks to one 
byte-sized entries in shadow space.
BBC re-arranges objects and aligns them to the base
of a block, to prevent metadata conflicts caused by 
multiple objects in one block. 
BBC pads each object to the next power of two,
so that each shadow table entry stores only 
the binary logarithm of the padded object size.
These significantly reduce the size of the shadow space, 
but perform \textit{approximate} bounds checking,
that tolerates pointers going out-of-bounds yet 
within the padded bound.  


\begin{figure}[t]
  \centering

  
  \begin{subfigure}[t]{0.4\columnwidth}
  \centering
 
      \begin{tikzpicture}[
        node distance=0pt,
        box/.style={
            node distance=0pt,outer sep=0pt,
            draw=black,
            fill=white, 
            font=\fontsize{8}{8}\ttfamily, 
            anchor=west
        },
        shadow/.style={
            node distance=0pt,outer sep=0pt,
            fill=palered,
            font=\fontsize{8}{8}\ttfamily, 
            anchor=west
        },
    ]
    \node[box=1.9cm, minimum width=1.6cm, 
            minimum height=0.8cm, thick](asan_obj) 
            at (-3.0, 2.0)
           {\textbf{OBJECT}};
    \node[shadow=1.9cm, minimum width=1.6cm, 
          minimum height=0.3cm, above=of asan_obj,
          yshift=0.05cm](asan_frag1)
          {}; 
    \node[shadow=1.9cm, minimum width=1.6cm, 
          minimum height=0.3cm, below=of asan_obj](asan_frag2) 
          {};
    \node[box=1.9cm, minimum width=1.6cm, 
          minimum height=0.4cm, thick,
          below=of asan_frag2](asan_obj2) 
           {\textbf{OBJECT}};
    \node[shadow=1.9cm, minimum width=1.6cm, 
          minimum height=0.3cm, below=of asan_obj2](asan_frag3) 
          {};
    \node[box=1.9cm, minimum width=1.6cm, thick, 
          minimum height=0.3cm, below=of asan_frag3](asan_ptr1) 
          {\textbf{P1}};
    \node[shadow=1.9cm, minimum width=1.6cm, 
          minimum height=0.2cm, below=of asan_ptr1](asan_frag4) 
          {};
   
    \draw[-, thick ] 
        ([xshift= 0.1cm] asan_ptr1.west) 
        -- 
        ([xshift = -0.2cm] asan_ptr1.west);

    \draw[-, thick] 
        ([xshift= -0.2cm] asan_ptr1.west) 
        -- 
        ([xshift= -0.2cm, yshift= -0.2cm] asan_obj.west);

    \path[arrow, thick] 
        ([xshift= -0.2cm, yshift= -0.2cm] asan_obj.west) 
        -- 
        ([yshift= -0.2cm] asan_obj.west);

    
    \node[box=1.9cm, minimum width=1.4cm, thick,
          minimum height=0.3cm](entry1)
          at (-1.0, 1.8)
          {};
   
    \node[shadow=1.9cm, minimum width=1.4cm, 
          minimum height=0.1cm,
          above=of entry1,
          yshift=0.02cm](shadow1)
          {};
    
    \node[shadow=1.9cm, minimum width=1.4cm, 
          minimum height=0.1cm, below=of entry1,
          yshift=-0.02cm](shadow2)
          {};

    \node[box=1.9cm, minimum width=1.4cm, thick,
          minimum height=0.1cm, below=of shadow2,
          yshift=-0.02cm](entry2)
          {};
    
    \node[shadow=1.9cm, minimum width=1.4cm, 
          minimum height=0.1cm, below=of entry2,
          yshift=-0.02cm](shadow3)
          {};
    
    \node[box=1.9cm, minimum width=1.4cm, 
          minimum height=0.1cm, below=of shadow3,
          draw=white,yshift=-0.02cm](shadow3)
          {\textbf{shadow space}};

    \draw[-, dotted] 
        ([yshift= 0.4cm] asan_obj.east) 
        -- 
        ([yshift= 0.15cm]entry1.west);
    
    \draw[-, gray, dotted] 
        ([yshift= -0.4cm] asan_obj.east) 
        -- 
        ([yshift= -0.2cm]entry1.west);

    \draw[-, gray, dotted] 
        ([yshift= 0.2cm] asan_obj2.east) 
        -- 
        ([yshift= 0.1cm]entry2.west);
    
    \draw[-, gray, dotted] 
        ([yshift= -0.1cm] asan_obj2.east) 
        -- 
        ([yshift= -0.1cm]entry2.west);

    \path[arrow, thick, gray] 
        ([yshift= -0.2cm] asan_obj.west) 
        -- 
        ([yshift= -0.2cm] asan_obj.east);
 
    \path[arrow, thick, gray] 
        ([yshift= -0.2cm] asan_obj.east) 
        -- 
        ([yshift= -0.1cm] entry1.west);

  \end{tikzpicture}
  \vspace*{-6mm}
  \caption{\small{Address Sanitizer}}\label{fig:asan}
  \end{subfigure}\hspace{2mm}
  ~ 
  \begin{subfigure}[t]{0.4\columnwidth}
  \centering
  \begin{tikzpicture}[
        node distance=0pt,
        box/.style={
            node distance=0pt,outer sep=0pt,
            draw=black,
            fill=white, 
            minimum height=0.3cm,
            font=\fontsize{8}{8}\ttfamily, 
            anchor=west
        },
        shadow/.style={
            node distance=0pt,outer sep=0pt,
            fill=palepurple,
            minimum height=0.3cm,
            font=\fontsize{8}{8}\ttfamily, 
            anchor=west
        },
    ]
    \node[box=1.9cm, minimum width=1.6cm, 
            minimum height=0.6cm, thick](mpx_obj) 
            at (-1.4, 2.0)
           {\textbf{OBJECT}};
    \node[shadow=1.9cm, minimum width=1.6cm, 
          minimum height=0.2cm, above=of mpx_obj,
          yshift=0.05cm](mpx_frag1)
          {}; 
    \node[shadow=1.9cm, minimum width=1.6cm, 
          minimum height=0.1cm, below=of mpx_obj](mpx_frag2) 
          {};
    \node[shadow=1.9cm, minimum width=1.6cm, 
          minimum height=0.1cm, below=of mpx_frag2](mpx_frag3) 
          {};
    \node[box=1.9cm, minimum width=1.6cm, thick, 
          minimum height=0.3cm, below=of mpx_frag3](mpx_ptr1) 
          {\textbf{P1}};
    \node[shadow=1.9cm, minimum width=1.6cm, 
          minimum height=0.2cm, below=of mpx_ptr1](mpx_frag4) 
          {};
    \node[box=1.9cm, minimum width=1.6cm, thick, 
          minimum height=0.3cm, below=of mpx_frag4](mpx_ptr2) 
          {\textbf{P2}};
    \node[shadow=1.9cm, minimum width=1.6cm, 
          minimum height=0.2cm, below=of mpx_ptr2](mpx_frag5) 
          {};
   
    \draw[-, thick ] 
        ([xshift= 0.1cm] mpx_ptr1.west) 
        -- 
        ([xshift = -0.2cm] mpx_ptr1.west);
    \draw[-, thick] 
        ([xshift= -0.2cm] mpx_ptr1.west) 
        -- 
        ([xshift= -0.2cm, yshift= -0.2cm] mpx_obj.west);
    \path[arrow, thick] 
        ([xshift= -0.2cm, yshift= -0.2cm] mpx_obj.west) 
        -- 
        ([yshift= -0.2cm] mpx_obj.west);

    \draw[-, thick] 
        ([xshift= 0.1cm] mpx_ptr2.west) 
        -- 
        ([xshift = -0.4cm] mpx_ptr2.west);
    \draw[-, thick] 
        ([xshift= -0.4cm] mpx_ptr2.west) 
        -- 
        ([xshift= -0.4cm, yshift= 0.2cm] mpx_obj.west);
    \path[arrow, thick] 
        ([xshift= -0.4cm, yshift= 0.2cm] mpx_obj.west) 
        -- 
        ([yshift= 0.2cm] mpx_obj.west);

    \node[box=1.9cm, minimum width=0.3cm, thick,
          minimum height=0.3cm](metad_lb)
          at (0.7, 2.0)
          {\textbf{LB}};
    \node[box=1.9cm, minimum width=0.3cm, thick,
          minimum height=0.3cm, right=of metad_lb](metad_ub)
          {\textbf{UB}};

    \node[box=1.9cm, minimum width=1.05cm, palemint, 
          minimum height=0.3cm, xshift=0.25cm,
          above=of metad_lb](metad_frag1)
          {};
    
    \node[box=1.9cm, minimum width=1.05cm, palemint,
          xshift=0.25cm,
          minimum height=0.3cm, below=of metad_lb](metad_frag1)
          {};

    \node[box=1.9cm, minimum width=0.3cm, thick,
          minimum height=0.3cm, below=of metad_frag1,
          xshift=-0.25cm](metad_lb2)
          {\textbf{LB}};
    \node[box=1.9cm, minimum width=0.3cm, thick,
          minimum height=0.3cm, right=of metad_lb2](metad_ub2)
          {\textbf{UB}};

    \node[box=1.9cm, minimum width=1.05cm, palemint, 
          minimum height=0.3cm, below=of metad_lb2,
          xshift=0.25cm](metad_frag2)
          {};
    \node[box=1.9cm, minimum width=1.05cm, draw=white, 
          minimum height=0.3cm, below=of metad_frag2]
          (metad_fraglabel)
          {\textbf{bounds table}};

    \draw[-, thick, gray] 
        ([xshift= -0.1cm] mpx_ptr1.east) 
        -- 
        ([xshift = 0.1cm] mpx_ptr1.east);
    \draw[-, thick, gray] 
        ([xshift= 0.1cm] mpx_ptr1.east) 
        -- 
        ([xshift= -0.3cm] metad_lb.west);
    \path[arrow, thick, gray] 
        ([xshift= -0.3cm] metad_lb.west) 
        -- 
        (metad_lb.west);

    \draw[-, thick, gray] 
        ([xshift= -0.1cm] mpx_ptr2.east) 
        -- 
        ([xshift = 0.3cm] mpx_ptr2.east);
    
    \draw[-, thick, gray] 
        ([xshift= 0.3cm] mpx_ptr2.east) 
        -- 
        ([xshift= -0.2cm] metad_lb2.west);
    
    \path[arrow, thick, gray] 
        ([xshift= -0.2cm] metad_lb2.west) 
        -- 
        (metad_lb2.west);

  \end{tikzpicture}
  \vspace*{-6mm}
  \caption{\small{MPX}}\label{fig:mpx}
  \end{subfigure}

\vspace*{1mm}
\caption{\small{Disjoint Metadata}}
\label{comparison}
\end{figure}
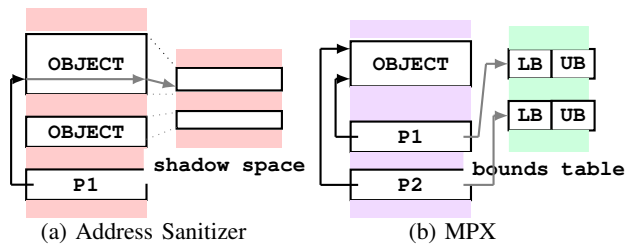

 
Address Sanitizer~\cite{address-sanitizer} 
(ASan) utilizes shadow space in a different way.
It pads each object with \textit{redzone}(rz) front and back
(Fig.~\ref{fig:asan}),
and considers access to this rz as out-of-bounds.
The errors (i.e.~access to rz) are identified by the value 
in the corresponding entry in the shadow.
At memory access, ASan derives the address of its 
corresponding entry from a pointer,
and the entry tells if the address is \textit{addressable} 
or not.     
ASan maps 8 bytes in application space into 
one byte in shadow space, and values in the bytes 
are written at object allocation. 
ASan also detects some dangling pointers, 
by forcing freed objects to stay in a so-called 
\textit{quarantine zone} for a while. 
Disadvantage of ASan is that its error detection relies 
on spatial or temporal \textit{distance}. 
It loses track of pointers going far beyond of rz and 
reaching another object's valid range, so  
fails to address false negatives caused by violation of  
\textit{intended referents}. 
The wider the rz, the more errors ASan detects.
\textit{use-after-free} errors cannot be detected, 
in the cases where dangling pointers are used to access objects 
after the pointer is freed from the quarantine. 
ASan detects most errors, but it is less 
deterministic in theory. 
    

Disjoint memory ranges can offer protection from metadata manipulation,
ranging from surrounding the memory area with unmapped pages
and randomizing its base address, to range checking all memory accesses. 

\subsection{Hardware-based Approaches}

Instruction set extensions for bounds checking~\cite{Devietti:2008:HAS:1353536.1346295, intel_mpx, 
Kwon13low-fatpointers:, SSM, 2665740} have
been proposed to overcome runtime overheads and 
limitations of software-based approaches.
Intel MPX~\cite{intel_mpx, DBLP:conf/snapl/NagarakatteMZ15, Oleksenko:2018:IME:3232754.3224423} 
is an ISA extension that provides a hardware-accelerated 
pointer checking using disjoint metadata.
MPX has four registers:  
one holds a pointer itself, two for upper and lower bounds,
and the last register keeps a copy of the pointer. 
If there is a mismatch between a pointer and the copy, 
MPX considers the pointer has been updated in un-instrumented 
code and gives up tracking of the pointer. 
This mechanism provides    
incremental deployment and seamless integration of codes.   
Reportedly, the MPX approach suffers due to lack of memory even with 
small working sets~\cite{Kuvaiskii:2017:SMS:3064176.3064192}
and has turned out to be slow for pointer-intensive programs, 
owing to the restricted number of special-purpose bounds 
registers (4 registers) is soon exceeded, requiring spill 
operations from regions of memory that themselves require 
management and consume D-cache bandwidth and capacity. 

\section{\pname{} Approach} 

In this section we provide a high-level description 
of how \pname{} handles per-object metadata 
efficiently. In a nutshell, the idea is to place
per-object metadata close to its object (normally 
in a header) and streamline metadata lookup 
by calculating the location from only
(1) an inbound pointer and (2) additional information 
tagged in the otherwise unused, top 16 bits.
We exploit the fact that
\textit{relative addresses} can be encoded in far
fewer bits than \textit{absolute} addresses provided
there is assistance from the memory manager to restrict
the distance between the allocation for an object and a
separate object for its metadata.  In many cases, the
metadata can be stored in front of the object, essentially
as a header, requiring only a single memory manager allocation.
The relative distance between object and metadata is then
normally sufficiently small to be encoded in relative form
in the top 16 bits.
But there are cases where relative location information
cannot be used, such as when allocating large objects
or with some small objects depending on their absolute address,
to avoid imposing alignment constraints on them.
We use a supplementary table only for these cases.
The top 16 bits encode when this is the case, and also
sufficient information to locate the supplementary entry.

We are now going to present the concept of \emph{frames}. 
In Section~\ref{sec:metadata-storage} we thoroughly
present how metadata are actually stored for each different object.


\begin{figure}
\noindent\resizebox{\columnwidth}{!}{%
\begin{tikzpicture}[
    node distance=0pt,outer sep=0pt,
    frame/.style={
        line width=1pt,
        draw=black,
        rounded corners=1mm,
        text width=#1,
        font={\sffamily\bfseries},
        align=center,
        text height=1pt,
        text depth=6pt
    },
    frame/.default=1.5cm,
    chosenframe/.style={
        line width=2pt,
        draw=black,
        fill=pastelteal,
        rounded corners=1mm,
        text width=#1,
        font={\sffamily\bfseries},
        align=center,
        text height=1pt,
        text depth=6pt
    },
    chosenframe/.default=1.5cm,
    membyte/.style={
        line width=1pt,
        draw=black,
        text width=#1,
        font={\sffamily\bfseries},
        align=center,
        text height=1pt,
        text depth=1pt
    },
    membyte/.default=1.5cm
]

\node[frame=0.30cm,label=left:$2^0$,draw] (1) {};
\node[frame=0.30cm,right=of 1] (2) {};
\node[frame=0.30cm,right=of 2] (3) {};
\node[frame=0.30cm,right=of 3] (4) {};
\node[frame=0.30cm,right=of 4] (5) {};
\node[frame=0.30cm,right=of 5] (6) {};
\node[frame=0.30cm,right=of 6] (7) {};
\node[frame=0.30cm,right=of 7] (8) {};
\node[frame=0.30cm,right=of 8] (9) {};
\node[frame=0.30cm,right=of 9] (10) {};
\node[frame=0.30cm,right=of 10] (11) {};
\node[frame=0.30cm,right=of 11] (12) {};
\node[frame=0.30cm,right=of 12] (13) {};
\node[frame=0.30cm,right=of 13] (14) {};
\node[frame=0.30cm,right=of 14] (15) {};
\node[frame=0.30cm,right=of 15] (16) {};

\widernode[chosenframe, label=left:$2^1$]{1}{2}{}{1_2}
\widernode[frame]{3}{4}{}{3_4}
\widernode[frame]{5}{6}{}{5_6}
\widernode[frame]{7}{8}{}{7_8}
\widernode[frame]{9}{10}{}{9_10}
\widernode[frame]{11}{12}{}{11_12}
\widernode[frame]{13}{14}{}{13_14}
\widernode[frame]{15}{16}{}{15_16}

\widernode[frame, label=left:$2^2$,draw]{1_2}{3_4}{}{1_4}
\widernode[frame]{5_6}{7_8}{}{5_8}
\widernode[frame]{9_10}{11_12}{}{9_12}
\widernode[frame]{13_14}{15_16}{}{13_16}

\widernode[frame,label=left:$2^3$,draw]{1_4}{5_8}{}{1_8}
\widernode[chosenframe]{9_12}{13_16}{}{9_16}

\widernode[chosenframe,label=left:$2^4$,draw]{1_8}{9_16}{}{1_16}

\node[membyte=0.30cm, fill=pastelteal, above=2 em of 1,
label=above:\texttt{a}] (byte1) {};
\node[membyte=0.30cm, fill=pastelteal, right=of byte1] (byte2) {};
\node[membyte=0.30cm, right=of byte2] (byte3) {};
\node[membyte=0.30cm, right=of byte3] (byte4) {};
\node[membyte=0.30cm, right=of byte4] (byte5) {};
\node[membyte=0.30cm, right=of byte5] (byte6) {};
\node[membyte=0.30cm, fill=pastelteal, right=of byte6,
label=above:\texttt{b}] (byte7) {};
\node[membyte=0.30cm, fill=pastelteal, right=of byte7] (byte8) {};
\node[membyte=0.30cm, fill=pastelteal, right=of byte8] (byte9) {};
\node[membyte=0.30cm, right=of byte9] (byte10) {};
\node[membyte=0.30cm, fill=pastelteal, right=of byte10,
label=above:\texttt{c}] (byte11) {};
\node[membyte=0.30cm, fill=pastelteal, right=of byte11] (byte12) {};
\node[membyte=0.30cm, fill=pastelteal, right=of byte12] (byte13) {};
\node[membyte=0.30cm, fill=pastelteal, right=of byte13] (byte14) {};
\node[membyte=0.30cm, right=of byte14] (byte15) {};
\node[membyte=0.30cm, right=of byte15] (byte16) {};
\node[membyte=0.30cm, left=of byte1, draw=none] {$\ldots$};
\node[membyte=0.30cm, right=of byte16, draw=none] {$\ldots$};

\node[above=0.1 em of byte13, xshift=1.5em, draw=none] (memlabel){Memory Space};
\node[above=0.1 em of 14, draw=none] (framelabel){Aligned Frames};

\path[arrow,ultra thick, gray] 
    
    (byte1) edge ([xshift=-2.7mm] 1_2.north)
    (byte2) edge ([xshift=2.7mm] 1_2.north)
    (byte7) edge ([xshift=-8.1mm] 1_16.north)
    (byte9) edge ([xshift=2.7mm] 1_16.north)
    (byte11) edge ([xshift=-8.1mm] 9_16.north)
    (byte14) edge ([xshift=8.1mm] 9_16.north)
;

\end{tikzpicture}
}
\caption{\small{Aligned frames in memory space}}
\label{fig:memory_frames}
\end{figure}
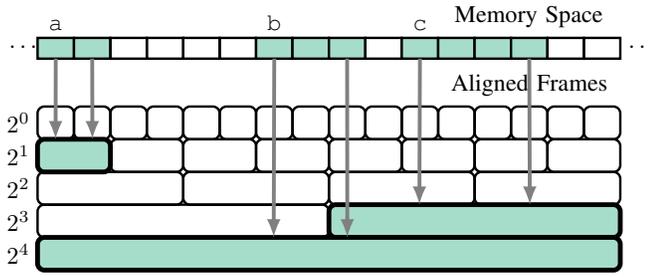

\subsection{Frame Definitions}

\pname{} stores metadata for all objects 
in separate blocks that are placed nearby by \pname{}'s
memory manager. User blocks are unchanged in layout, but their
metadata can be accessed with minimal additional cache misses.
We record a mapping between a block and its metadata using
the top 16 bits of a 64-bit pointer, which are spare in contemporary CPUs. 
As we show below, the code to resolve the metadata address 
from the inbound pointer is very fast.
It does not involve any time-consuming traversal/lookup for metadata access.


To record the relative offset between a block and its metadata
we define a logical structure over the whole
data space of a process, including statics, stack, and heap. 
The \pname{} structures are based on the concept of \textit{frames},
defined as memory blocks
that are $2^n$-sized and aligned by their size,
where $n$ is a non-negative integer.
A frame of size $2^n$ is called \textit{n-frame}. 
A memory object \textit{x} will intrinsically lie 
inside at least one bounding frame, 
and \textit{x}'s \textit{wrapper frame} is defined 
as the smallest frame completely containing $x$ inside.
For instance, in Fig.~\ref{fig:memory_frames}, 
objects \texttt{a,b} and \texttt{c}'s wrapper frames
are $(n=1)$-frame (or 1-frame), 4-frame, and 3-frame, 
respectively. 
For $0\leq$m$<$n, 
we call $m$-frames placed inside a $n$-frame 
\texttt{f}, \texttt{f}'s \textit{subframes}.


Frames have several interesting properties. 
Firstly, a $n$-frame is aligned by $2^m$ 
for all $m<n$. 
Secondly, an object's wrapper frame size need not 
grow in proportion to the object's size. As shown 
in Fig.~\ref{fig:memory_frames}, the object 
\texttt{b} has a larger wrapper frame than \texttt{c},
even though \texttt{b}'s size is smaller. This is because 
the wrapper frame size for an object is determined 
by both the object's size and location.   
Thirdly, a memory object \textit{x}'s wrapper frame 
is a frame containing \textit{x} in it, and having
the base (i.e.~lower bound) and upper bound of \textit{x} 
in its lower-addressed $(n-1)$-subframe and 
higher-addressed $(n-1)$-frame, respectively. 
It is trivial to prove this
as presented in Appendix~\ref{appendix1}. 

Following basic {\tt malloc} semantics, \pname{} 
does not natively support object movement or growth
(we reset its wrapper frame at {\tt realloc}).
Therefore, there exists the unique wrapper frame 
for each object, and it is determined 
at memory allocation. We use it as a reference point 
to obtain relative location information for each object, 
since it does not change 
during the life time of the object. 
At memory allocation, we determine the wrapper frame 
for the allocated object and tag the relative 
location in the pointer using the wrapper frame size.

\subsection{Frame Selection}\label{sec:frame_selection}

Now we show how to get the size of the wrapper frame,
given an object. 
We call an object whose wrapper frame is $n$-frame 
an \textit{n-object}. 
For any $k$-object $o$,
its wrapper frame (i.e., $k$-frame) is aligned by 
$2^k$ by definition, addresses of all the bytes in the 
frame have the same value setting of most significant 
$(64-k)$ bits, and so do the addresses of all bytes in $o$. 
In addition, the base and upper bound are located in the 
lower and higher-addressed $(k-1)$-frame, respectively. 
This means that the  $(k-1)_{th}$ least significant bit 
of the base and upper bound must be negated. 

Based on these, we can get $k$, $\log_2$ of 
the size of $o$'s wrapper frame.
Let's say $(b_{63},..., b_{1},b_{0})$ and 
$(e_{63},..., e_{1},e_{0})$ are bit vectors 
of $k$-object $o$'s base and upper bound 
respectively, and $X$ is a \textit{don't care} value.
We get the $\log_2$(wrapper frame size) 
by performing \texttt{XOR} (exclusive OR) and 
\texttt{CLZL} (count leading zeros) operations 
as follows ($b_{63}$ is the most significant):

\begin{equation*}
\begin{adjustbox}{max width=210pt,margin=-5pt -3pt -5pt -18pt}
$
\begin{array}{llllllll}
(b_{63},&...,&b_{k},&b_{(k-1)},&b_{(k-2)},&...,&b_{0})& \\
(e_{63},&...,&e_{k},&e_{(k-1)},&e_{(k-2)},&...,&e_{0})&{\tt XOR}\\
\hline
(0,&...,&0,&1,&{\tt X},&...,&{\tt X})&{\tt CLZL}\\
\hline
&&&&&&(64-k)& \\
\hline
\end{array}
$
\end{adjustbox}
\end{equation*}

We then get $k$ by subtracting the result of {\tt clzl} operation 
from 64: $64 - (64-k) = k$.

\subsection{Metadata Storage Management}
\label{sec:metadata-storage}



As said, \pname{} stores metadata per object at an address 
that can be derived from a tagged pointer. 
Objects carry their metadata in a \textit{header} 
associated with themselves.
We encode the relative location of metadata using 
the wrapper frame size of each object.
Assuming the header has a size of \emph{h}, the base 
address of the object is the header address plus 
\emph{h} bytes,
hence, the base of an object does not need to be
stored for bounds checking. 
For instance, in Fig.~\ref{fig:get_metadata},
{\tt a, b} and {\tt c} are all objects containing 
a \emph{h}-sized header. 




\pname{} considers two core types of objects, depending
on their wrapper frame size, namely \emph{small-framed} 
and \emph{big-framed} objects, and
re-structures the virtual address space as follows.
\pname{} divides user space into
\textit{slots} with a fixed size of $2^{15}$ bytes
and aligned to their size, i.e., $(N=15)$-frames.  
Slots are set to a size of $2^{15}$ so that the offset 
to the header of small-framed objects ($(N\leq15)$-objects)  
can be encoded in the unused 15 bits of a pointer. 
In Fig.~\ref{fig:get_metadata}, 
$d_a$ is the offset to the header of the small-framed 
object {\tt a}. 
Typically, we expect the number of big-framed objects 
to be low compared to small-framed objects.

\begin{figure}[t] 
\noindent\resizebox{0.8\columnwidth}{!}{%
\centering
\begin{tikzpicture}[
    node distance=0pt,
    bits/.style={
        line width=1pt,
        draw=black,
        text width=#1,
        font={\sffamily\bfseries},
        align=center,
        text height=3pt,
        text depth=1pt,
        font=\fontsize{6}{6}\ttfamily, 
        anchor=west
    }
]
    \def\mynodedistance{1cm}
    \def\txtdist{0.3cm}

    \node[bits=0.4cm, fill=pastelteal, thick, anchor=north west] 
        at (1, -2.5) (flag) {flag};
    \node[bits=1cm, thick, fill=pastelteal, right=of flag] (tag) {tag};
    \node[bits=3cm, thick, right=of tag] (addr) {address};

    \node[bits=0.2cm, draw=white, below=of flag, align=left] (myflag1)
        {1};
    \node[bits=0.2cm, draw=white, below=of flag, yshift=-0.3cm, align=left] (myflag2) 
        {0};
    
    \node[bits=0.3cm, draw=white, below=of tag] (mytag1)
        {offset};
    \node[bits=0.3cm, draw=white, below=of tag, yshift=-0.3cm] (mytag2) 
        {N};
    
    \node[bits=2cm, draw=white, below=of addr, align=left,
          xshift=-0.5cm] (mytag1)
        {if N$<=$15};
    \node[bits=3cm, draw=white, below=of addr, yshift=-0.3cm, align=left] (mytag2) 
        {if N$>$15};
    
    \draw[|<->|] ([yshift=\txtdist] addr.west) -- node[above] 
    {\fontsize{6}{6}\scriptsize\ttfamily 48} ([yshift=\txtdist] addr.east);
    \draw[|<->|] ([yshift=\txtdist] tag.west) -- node[above] 
    {\scriptsize\ttfamily 15} ([yshift=\txtdist] tag.east);
    \draw[|<->|] ([yshift=\txtdist] flag.west) -- node[above] 
    {\scriptsize\ttfamily 1} ([yshift=\txtdist] flag.east);
    
\end{tikzpicture}
}
\caption{\small{Tagged Pointer:
the tag depends on the value of N (binary logarithm of 
the wrapper frame size of a referent object).}}
\label{fig:tagged_pointer}
\end{figure}
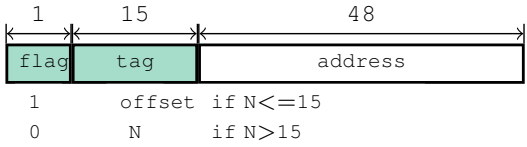

One extra bit, in particular the most significant, 
is taken for the \textit{flag} property, which indicates if the object is 
small-framed or big-framed ($(N>15)$-objects)
as shown in Fig. \ref{fig:tagged_pointer}.

The descriptor of big-framed objects requires more bits 
of information (described in~\ref{sssec:bigframedobj}).
For big-framed objects, \pname{}   
creates one array -- can be interpreted as \emph{shadow space} 
-- that holds additional location information. The corresponding 
entry of a big-framed object is then directly accessed only 
with a tagged pointer. We stress here that this array is 
not needed for lookups associated with small-framed 
objects, and is smaller than typical shadow memory implementations
where each entry corresponds to an aligned memory word.

\subsubsection{Small-framed Objects}\label{sec:small_frame} 

Since small-framed objects  
are placed in a single slot, we simply tag a 
pointer with the \textit{offset} from the base 
address of the slot to the header of the object.
We further turn on the most significant bit of the pointer
to indicate that the particular object is small-framed. 
When we retrieve metadata from a header of an small-framed
object, (i.e., flag=={\tt 1}) inbound 
(in-slot) pointers are derived to the base 
of the slot by zeroing the least significant 
$15$ ($log_2 (slot\_size)$) bits, and then to the address 
of the header by adding the offset to the base address 
of the slot as follows:

\begin{lstlisting}
FLAG_MASK = ~(1ULL << 63);
offset = (tagged_ptr & FLAG_MASK) >> 48; 
slotbase = untagged_ptr & (~0ULL << 15);
header_addr = slotbase + offset;
obj_base = header_addr + header_size;
\end{lstlisting}

\subsubsection{Big-framed Objects}\label{sec:big_frame} 
\label{sssec:bigframedobj}

Big-framed objects span several slots, thus their offset 
cannot be solely used as their relative location. 
Zeroing the least 15 significant bits (log slotsize)
of a pointer does not always lead to a unique slot base.
In Fig.~\ref{fig:get_metadata}, 
an object {\tt b}'s inbound pointer can derive 
two different slots ({\tt slot0} and {\tt slot1}) 
depending on the pointer's value, and that is the case 
for object {\tt c} ({\tt slot1} and {\tt slot2}).
Hence, for big-framed objects, we need to store additional 
location information in our supplementary table. 

During program initialisation, we create 
an array, and each entry is mapped
to a 16-frame. We call such a frame a 
\textit{division}. Each entry contains one sub-array
and the sub-array per division is called
a \textit{division array}. Each division array
contains the fixed number of entries in the current
implementation as follows: 

\begin{lstlisting}
typedef struct Entries {
  void *header_addr; /* The header address */ 
} EntryT; 

typedef struct ShadowTableEntries {
  EntryT division_array[48]; /* 64-16 */  
} DivisionT; 
\end{lstlisting}

\tikzstyle{arrow}=[draw, -latex] 
  
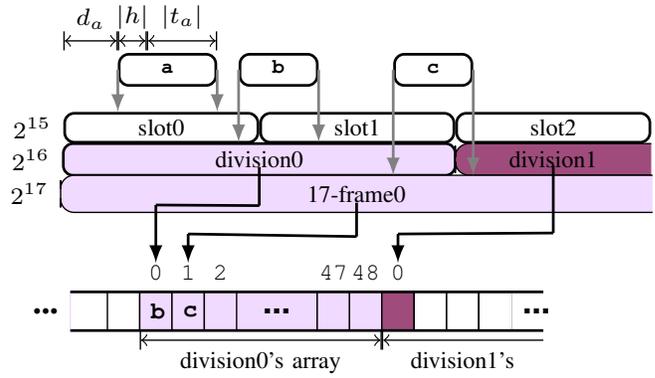
\begin{figure}
\noindent\resizebox{\columnwidth}{!}{%
\begin{tikzpicture}[
    node distance=0pt,
    object/.style={
        rounded corners=1mm,
        draw=black,
        fill=white, 
        minimum width=1.6cm,
        text height=2.0pt,
        text depth=0.5pt,
        font=\fontsize{6}{6}\ttfamily, 
        anchor=west
    },
    frame/.style={
        rounded corners=1mm,
        draw=black,
        thick,
        font=\scriptsize,
        minimum width=2cm,
        minimum height=0.3cm,
        inner sep=0pt
    },
    rightflat/.style={
        append after command={%
            \pgfextra
            \fill[fill=#1] 
            (\tikzlastnode.south east) 
            [rounded corners] -| 
            (\tikzlastnode.west) |- 
            (\tikzlastnode.north) 
            [sharp corners] -| cycle;
        \draw[rounded corners] 
            (\tikzlastnode.south east) -| 
            (\tikzlastnode.west) |- 
            (\tikzlastnode.north east);
            \endpgfextra},
        minimum width=2cm,
        inner sep=1.8pt,
        font=\scriptsize,
        thick
    },
    leftflat/.style={
        append after command={%
            \pgfextra
            \fill[fill=#1] 
            (\tikzlastnode.north west) 
            [rounded corners] -| 
            (\tikzlastnode.east) 
            |- (\tikzlastnode.south) 
            [sharp corners] -| cycle;
        \draw[rounded corners] 
            (\tikzlastnode.north west) -| 
            (\tikzlastnode.east) |- 
            (\tikzlastnode.south west);
        \endpgfextra},
        minimum width=2cm,
        text height=2.0pt,
        text depth=0.5pt,
        fill=bblue,
        thick
    },
    entry/.style={
        node distance=0pt,outer sep=0pt,
        draw=black,
        text width=#1,
        font={\sffamily\bfseries},
        align=center,
        minimum width=0.1cm,
        minimum height=0.4cm
    },
]
    \def\txtdist{0.35cm}
    \node[frame, thick,  
        label=left:{\scriptsize\ttfamily $2^{15}$}] 
        at (0,1.5) (1) {slot0};
    \node[frame,right=of 1] (2) 
        {slot1};
    \node[frame,right=of 2] (3) 
        {slot2};
    \widernode[frame, fill=palepurple, 
        label=left:{\scriptsize\ttfamily $2^{16}$}]
        {1}{2}{\scriptsize division0}{1_2};
    \node[rightflat=ppurple, below=of 3, thick] (4) 
        {\scriptsize division1};
    \widernode[rightflat=palepurple, 
        label=left:{\scriptsize\ttfamily $2^{17}$}]
        {1_2}{4}{\scriptsize 17-frame0}{1_4};

    \node[entry=0.15cm, draw= palepurple, label=left:{\textbf{...}}
         ]
        at (-0.75,-0.4)  (entry0) {};
    \node[entry=0.10cm, right=of entry0] (entry1) {};
    \node[entry=0.10cm, fill=palepurple, right=of entry1,
        label=above:{\scriptsize\ttfamily 0}] (entry2) {\scriptsize\ttfamily b};
    \node[entry=0.10cm, fill=palepurple, right=of entry2,
        label=above:{\scriptsize\ttfamily 1}] (entry3) {\scriptsize\ttfamily c};
    \node[entry=0.10cm, fill=palepurple, right=of entry3,
        label=above:{\scriptsize\ttfamily 2}] (entry4) {};
    \node[entry=0.60cm, fill=palepurple, right=of entry4] (entry5) {...};
    \node[entry=0.10cm, right=of entry5, fill=palepurple,
        label=above:{\scriptsize\ttfamily 47}] (entry6) {};
    \node[entry=0.10cm, right=of entry6, fill=palepurple,
        label=above:{\scriptsize\ttfamily 48}] (entry7) {};
    \node[entry=0.10cm, right=of entry7, fill=ppurple,
        label=above:{\scriptsize\ttfamily 0}] (entry8) {};
    \node[entry=0.10cm, right=of entry8] (entry9) {};
    \node[entry=0.10cm, right=of entry9] (entry10) {};
    \node[entry=0.10cm, right=of entry10] (entry11) {};
    \node[entry=0.10cm, draw=white, right=of entry11] 
        (entry12) {\textbf{...}};
    \draw[-,thick] (entry0.north west) -- (entry12.north east);
    \draw[-,thick] (entry0.south west) -- (entry12.south east);
   
    \node (obj_a) [object, thick, minimum width=1cm] at (-0.45, 2.1) 
        {\textbf{a}};
    \node (obj_b) [object, thick, minimum width=0.8cm] at (0.8, 2.1) 
        {\textbf{b}};
    \node (obj_c) [object, thick, minimum width=0.8cm] at (2.4,2.1) {\textbf{c}};

    \path[arrow, thick, gray] (obj_a.west) -- (obj_a.west|-3.north);
    \draw[arrow, thick, gray] (obj_a.east) -- (obj_a.east|-3.north);
    
    \draw[arrow, thick, gray] (obj_b.west) -- (obj_b.west|-1_2.north);
    \draw[arrow, thick, gray] (obj_b.east) -- (obj_b.east|-1_2.north);
    
    \draw[arrow, thick, gray] (obj_c.west) -- (obj_c.west|-1_4.north);
    \draw[arrow, thick, gray] (obj_c.east) -- (obj_c.east|-1_4.north);
   
    \draw[-, thick] ([yshift= 0.1cm] 1_2.south) 
              -- 
             ([yshift= -0.3cm] 1_2.south);
    
    \draw[-,thick] ([yshift= -0.3cm] 1_2.south) 
              -- 
             ([yshift= 0.9cm] entry2.north);
    
    \path[arrow,thick] ([yshift= 0.9cm] entry2.north) 
             -- 
             ([yshift= 0.3cm] entry2.north);
   
   \draw[-, thick] ([yshift= 0.1cm] 1_4.south) 
              -- 
             ([yshift= -0.25cm] 1_4.south);
    
    \draw[-,thick] ([yshift= -0.25cm] 1_4.south) 
              -- 
             ([yshift= 0.6cm] entry3.north);
    
    \path[arrow,thick] ([yshift= 0.6cm] entry3.north) 
             -- 
         ([yshift= 0.3cm] entry3.north);

    \draw[-, thick] ([yshift= 0.1cm] 4.south) 
              -- 
             ([yshift= -0.6cm] 4.south);
    
    \draw[-,thick] ([yshift= -0.6cm] 4.south) 
              -- 
             ([yshift= 0.6cm] entry8.north);
    
    \path[arrow,thick] ([yshift= 0.6cm] entry8.north) 
             -- 
         ([yshift= 0.3cm] entry8.north);

    \draw[|<->|] ([yshift=0.3cm] obj_a.west) -- node[above] 
        {\scriptsize\ttfamily $d_a$} ([yshift=0.9cm]1.west);
    
    \draw[|<->|] ([yshift=0.3cm] obj_a.west) 
        -- node[above] {\scriptsize\ttfamily $|h|$} 
        ([yshift=0.3cm, xshift=0.3cm] obj_a.west);
    
    \draw[|<->|] ([yshift=0.3cm,xshift=0.3cm] obj_a.west) 
        -- node[above] {\scriptsize\ttfamily $|t_a|$} 
        ([yshift=0.3cm] obj_a.east);
    
    \draw[|<->|] ([yshift=-0.3cm] entry2.west) 
        -- node[below] 
        {\scriptsize division0's array} 
        ([yshift=-0.3cm] entry7.east);

    \draw[|<-] ([yshift=-0.3cm] entry8.west) 
        -- node[below] 
        {\scriptsize division1's} 
        ([yshift=-0.3cm] entry12.east);
     
    \end{tikzpicture}
}
\caption{{\small Access to division array:  
the object {\tt a} is small-framed, while 
{\tt b} and {\tt c} are big-framed. $d_a$ 
is the offset to {\tt a}. 
$h$ denotes a header and $|t_a|$ is the size of {\tt a}.
{\tt b} and {\tt c}'s entries are mapped to the same 
division array. The entries in the division arrays 
store their corresponding object's header location,
while the small-framed object {\tt a} does not have an entry.
Only one entry of division1's array
is actually used, since the division is not aligned by 
$2^{17}$.}}
\label{fig:get_metadata}
\end{figure}

Contrary to small-framed objects, we tag binary logarithm 
of their wrapper frame size (i.e.~$N$==$log_2 2^N$)
in pointers to big-framed objects.
The address of an entry in a division array is then 
calculated from an inbound pointer and the $N$ value,
and the entry holds the address of a header.
By definition, a wrapper frame of an $(N \geq 16)$-object 
is aligned by its size, $2^N$, therefore, the 
frame is also aligned by $2^{16}$. 
This implies that a $(N \geq 16)$-frame shares
the base address with certain division, and is  
mapped to one division. In addition, each object 
has an intrinsic $N$-value, since each object 
has one wrapper frame. Each $2^{16}$ frame is mapped 
to one division array, so we keep the header address 
of each $(N \geq 16)$-object in one of entries of 
the division array.

 
Each $(N \geq 16)$-object maps to one division
array, but that division array contains entries for 
multiple big-framed objects.
In Fig. \ref{fig:get_metadata}, both \texttt{division0} 
and \texttt{17-frame0} are mapped to \texttt{division0}.
Their mapped division (\texttt{division0}) is aligned 
by $2^{17}$ at minimum, while \texttt{division1} 
is aligned by $2^{16}$ at max.


Here, the tag $N$ is used 
as an index to identify big-framed objects mapped 
to the same division array.  
For each $N \geq 16$, at most one $N$-object is 
mapped to one division array, and the proof is 
presented in Appendix~\ref{appendix2}. 
We use the value $N$ as an index
of a division array, and tag $N$ in the pointer.
Given a $N$ value-tagged pointer ({\tt flag==0}), 
we derive the address of an entry as follows:

\begin{lstlisting}
/* Ubase: division base of userspace's base
   scale: binary log (division_size), i.e. 16
   TABLE: address of a supplementary table */
/* p is assumed untagged here */
framebase = p & (~0ULL << N); 
TABLE_index = (framebase - Ubase) / (1ULL << scale);
DivisionT *M = TABLE + TABLE_index; 
EntryT *m = M->division_array;
EntryT *myentry = m + (N - scale);
header_addr = myentry->header_addr; 
\end{lstlisting}


%

The base of the wrapper frame (i.e.~the base of 
the division) is obtained by zeroing the least 
significant $N$ bits of the pointer.
We, then, get the address to its division array 
from the distance from the base of virtual address 
space and $log_2 (division\_size)$ ($2^{16}$).
Finally we access the corresponding entry with the
index $N$ in the division array.  



Entries in a division array may not always be used, 
since an entry corresponds to one 
big-framed object, which is not necessarily 
allocated at any given time, 
e.g.\ if object {\tt b} is not allocated in the space
in Fig.~\ref{fig:get_metadata}, 0th element of 
{\tt division0}'s array would be empty.
This feature is used for detecting some dangling
pointers, and more details are explained in 
section~\ref{sec:dangling}. 

Unlike existing approaches using shadow space, \pname{} 
does not re-align objects to avoid conflicts in entries.
Our \textit{wrapper frame-to-entry} mapping 
allows wrapper frames to be overlapped, 
that gives full flexibility to memory manager.

We could use different forms of a header such as a 
\textit{remote} header or a \textit{shared} header 
for multiple objects, with considering a cache line, 
stack frame, or page.
In addition, although we fixed the division size ($2^{16}$),
future designs may offer better flexibility in size,
as long as entries are not overlapped.

We showed how to directly access per-object 
metadata only with a tagged pointer. 
Our approach eliminates expensive traversal in the
data structure; gives great flexibility to associate metadata
with each object; gives full freedom to arrange 
objects in memory space, that removes re-alignment
of objects unlike existing approaches using shadow space. 
This mechanism can be exploited for other purposes:
the metadata can hold any per-object data. 

\section{\pname{} Implementation}

This section describes the current implementation 
of \pname{} 
which is largely built using LLVM.
Additionally,  we discuss 
how we offer compatibility with existing code.


\subsection{Overview}

\tikzstyle{arrow}=[draw, -latex] 

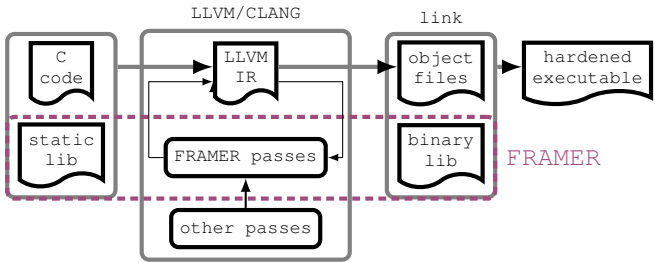
\begin{figure}[t]
  \resizebox{\columnwidth}{!}{
  \begin{tikzpicture}
    
    \node[docushape, minimum width=0.4cm] at (-4,0) (c_src) {C\\ code};  
    \node[docushape, minimum width=0.4cm, below=of c_src, yshift=0.6cm] (staticlib) {static\\ lib};  
    \node[docushape, minimum width=0.4cm, right=of c_src,
        xshift=0.6cm] (llvmIR) {LLVM\\ IR};  
    \node[docushape, minimum width=0.4cm, right=of llvmIR,
         xshift=0.6cm] (objfile) {object\\ files};  
    \node[docushape, minimum width=0.4cm, below=of objfile, 
       yshift=0.6cm] (binarylib) {binary\\ lib};  
    \node[docushape, minimum width=0.4cm, right=of objfile,
        xshift=-0.5cm] (exe) {hardened\\ executable};  

    \node[mymodule, minimum width=0.4cm, below=of llvmIR, yshift=0.4cm] 
        (framerpass) {\pname{} passes};
    \node[mymodule, minimum width=0.4cm, below=of framerpass,
        yshift=0.6cm] (otherpass) 
        {other passes};
  
    \node[myshade, fit=(c_src) (staticlib),
            minimum width=1cm, minimum height=2.2cm,
            yshift=-0.1cm] 
        (input) {};
        \node[myshade, fit=(binarylib) (objfile),
            minimum width=1cm, minimum height=2.2cm,
            label=above:{\scriptsize\ttfamily link},
            yshift=-0.1cm] 
        (output) {};
    \node[myshade, 
            fit=(llvmIR) (framerpass) (otherpass),minimum width=2.8cm,
            label=above:{\scriptsize\ttfamily LLVM/CLANG}] 
        (llvm) {};
    \node[myshade, draw=ppurple, densely dashed,  
        fit=(staticlib) (framerpass) (binarylib), 
            minimum width=4.5cm, minimum height=1.1cm,
            yshift=-0.1cm,
            label=right:{\fontsize{10}{10}\color{ppurple}\ttfamily \pname{}}] 
        (framermodule) {};

    \draw[-, to path={-| (\tikztotarget)}] 
        ([yshift=-0.2cm]llvmIR.east) edge ([xshift=0.2cm] framerpass.east);
    \path[arrow] ([xshift=0.2cm] framerpass.east) -- (framerpass.east);

    \draw[latex-, to path={-| (\tikztotarget)}] 
        ([yshift=-0.2cm] llvmIR.west) edge ([xshift=-0.2cm] framerpass.west);
    \draw[-] ([xshift=-0.2cm] framerpass.west) -- (framerpass.west); 
    
    \draw[-latex, draw=gray, ultra thick] 
        (input.east|-llvmIR.west) -- (llvmIR.west); 
    
    \draw[-latex, draw=gray, ultra thick] 
        (llvmIR) -- (objfile); 
    
    \draw[-latex, thick] 
        (otherpass) -- (framerpass); 
    
    \draw[-latex, draw=gray, ultra thick] 
        (output.east|-exe.west) -- (exe.west); 
      
  \end{tikzpicture}
 }   
  \caption{\small{Overall architecture of \pname{}}}
  \label{fig:overall_architecture}
\end{figure}



There are three main parts to our implementation: 
\pname{} LLVM passes, and the static library (lib), 
and the binary lib 
in the dashed-lined box as shown in Fig.~\ref{fig:overall_architecture}.
The target C source code and our hooks' functions
in the static lib are first compiled to LLVM intermediate
representation (IR).
Our main transformation pass 
instruments memory allocation, deallocation, memory access,
or optionally pointer arithmetic in the target code in IR. 
In general, instrumentation 
simply inserts a call to lib functions,
performing metadata update or bounds checking,
however,
our using of header-attached objects and tagged pointers requires 
extra program transformation at compile-time. 
The third part is wrappers around {\tt malloc} family
routines and string functions.
Compiler optimizations 
are discussed in Section~\ref{sec:optimisation}.

The flexible structure of LLVM allowed these to be implement 
using function interposition and two additional IR traversals, 
but we also had to modify the LLVM framework slightly.
Our main transformation is implemented as a LLVM Link Time 
Optimization (LTO) pass for whole program analysis, and run 
as a LTO pass on gold linker \cite{llvmGoldLinker}, however, 
incremental compilation is also possible. 


We also insert a prologue that is performed on
program startup. The prologue reserves 
address space for the supplementary metadata table, 
and pages are allocated on demand.


\subsection{Transformation of Memory Allocation}\label{sec:transform_alloc}


We instrument memory allocation and deallocation to update
metadata, and also transform the target IR code at compile time
for a header and tagged pointer.


\subsubsection{Stack-allocated Objects (address-taken locals)}

In implementation of a header-attached object, 
\pname{}'s transformation pass creates a new object 
in a structure type, that has 
two fields: a header's type and the same type as the 
original allocation as shown below:

\begin{lstlisting}
struct __attribute__((packed)) newTy {
  HeaderTy hd; 
  Ty obj; /* Ty is an original object's type */
};
\end{lstlisting} 

It then inserts a callsite to our hook function at allocation. 
The hook decides if it is small or big-framed, 
updates metadata in the header, and also 
in the entry for big-framed objects. It then creates a 
tag (offset or N value), and moves the pointer to the
second field whose type is the actual allocated type
in the target program. The hook returns a
tagged pointer. The allocation of an original object 
is removed by \pname{}'s pass, after the 
pass replaces all the pointers to the original object    
with the tagged pointer to the new object.

We instrument function epilogues to reset entries for 
big-framed non-static objects. 
Currently we instrument all the epilogues,
but this instrumentation can be removed 
for better performance. 


\subsubsection{Statically-allocated objects (address-taken globals)}\label{global_object_transformation}

Transformation on static objects (global variables) 
is similar to handling local ones on the stack.  
Creating a new global object with a header attached is 
straightforward, however, other parts of implementation  
are more challenging. 

Some operations on stack objects (decision of a wrapper 
frame size, creating a tag, updating metadata, and 
tagging a pointer) are performed in the hook, 
and the role of the transformation
pass is to replace all the pointers 
to an original object with the return value 
of the hook (i.e.~a tagged one). 
This cannot be applied to global objects, since the return
value of a function is non-constant, while the original pointer's
users having it as an operand may be constant. For example,
an original pointer (compile-time constant),
may be an initializer of other global/static objects, 
or an operand of some {\tt constant expression} (LLVM ConstExpr)~\cite{llvmConstantExpr}. 
Global variables' initializer and ConstExpr's 
operands must be constant,  
hence, 
the operations performed in a hook
for stack objects should be done by a transformation pass
for global objects.

In addition, while the tag should be generated at compile-time,
the wrapper frame size is determined by their 
actual addresses on memory, that are known only at run-time. 
To implement a tagged pointer 
generated from run-time information at compile-time, 
\pname{}'s transformation pass builds ConstExpr 
of (1) the wrapper frame size $N$ 
(2) offset, 
(3) tag and flag selection 
depending on its wrapper frame size, 
(4) pointer arithmetic operation
to move the pointer to the second field, and then finally 
(5) constructs a tagged pointer based on them.
The original pointers are replaced 
with this constant tagged pointer. The concrete value of 
the tagged pointer is then propagated at run-time, when 
the memory addresses for the base and bound are assigned.

\pname{} inserts a call to a function at the entry of main
function for each object. The function updates metadata 
in the header and the address in the entry, for
big-framed ones, during program initialisation.



   

\subsubsection{Heap objects}

We interpose calls to {\tt malloc}, {\tt realloc}, 
and {\tt calloc} at link time with wrapper functions 
in our binary libraries. 
The wrappers add the user-defined size by the 
header size; call {\tt malloc} and {\tt realloc};
and the rest of operations are the same as the hook 
for stack objects.
{\tt calloc} takes the number of elements and the size 
of an element, so we add minimum number of elements to 
hold the header (This allows spare bytes at the end 
of the object). 

We also interpose {\tt free} with our wrapper. 
This performs resetting an entry for a big-framed object, 
and releasing the object with the hidden base 
(i.e.~the address of the header).

\subsection {Memory Access}\label{sec:implementation_mem_access}


\pname{}'s transformation pass 
inserts a call to our bounds checking function 
right before each {\tt store} and {\tt load}, such that 
each pointer is examined and its tag stripped-off 
before being dereferenced.
The hook extracts the tag from a pointer, 
gets the header location,
performs the check using metadata in the header, 
and then returns an untagged pointer after cleaning 
the tag. The transformation pass replaces a tagged pointer 
operand of {\tt store}/{\tt load} with an untagged one
to avoid segmentation fault caused by dereferencing it.

Bounds checking and tag cleaning are also performed
on {\tt memcpy}, {\tt memmove} and {\tt memset} 
in similar way. (Note that 
LLVM overrides the C lib functions to their intrinsic 
functions~\cite{llvmMemFunc}). 
{\tt memmove} and {\tt memcpy} has two pointer operands, 
so we instrument each argument separately. 

As for string functions, we interpose these at link time. 
Wrapper functions performs checking
on strings, call {\tt real} functions with pointers
with a tag removed, and then restore tagged pointers.  


\subsection{Interoperability}

\pname{} ensures \textit{compatibility} 
between instrumented 
modules and regular pointer representation in pre-compiled
non-instrumented
libraries. We strip off tagged pointers before passing 
them to non-instrumented functions. 

\pname{} adds a header to objects for tracking, but this 
does not introduce incompatibility, since 
it does not change the internal memory layout of objects 
or pointers.



\section{\pname{} Applications}

In this section we discuss how \pname{} can be used 
for building security applications. We focus mainly 
on spatial safety. Nevertheless, we discuss additional 
case studies related to temporal safety and type checking.


\subsection{Spatial Memory Safety}

In a nutshell, \pname{} tracks individual memory allocations, 
and associates metadata with them. The metadata is
stored in the header associated with an object, 
and the offset, or the wrapper frame information ($N$ value),
is tagged in the pointer. We update the metadata 
and tag at object allocation; metadata is retrieved
at memory access ({\tt store}, {\tt load} and 
selected standard library functions). The tagged pointers must be
stripped of their tag before being dereferenced to
prevent segmentation fault. 
Unlike other object-tracking or relative location-based 
approaches, \pname{}
can tackle legitimate pointers outside the object bounds
without padding objects or requiring metadata retrieval 
or bounds checking at pointer arithmetic operations.


In this section, we describe how \pname{} performs 
bounds checking at run-time. 

\begin{table}[t]
{\small
\hfill{}
\resizebox{\columnwidth}{!}{
\begin{tabular}{|l|l|l|}       
\hline
& Original C & Instrumented C \\ 
\hline
\hline

\rowcolor{gray!20} 


\rowcolor{gray!20} 
1 & 
& 
\texttt{struct newTy\{HeaderTy hd;\textbf{int A[10];}\}};\\


2 &
\texttt{int A[10];} & 
\texttt{struct newTy new\_A;} \\ 

\rowcolor{gray!20} 
3 & 
& 
\texttt{tagged = handle\_alloc(\&new\_A, A\_size);}\\

\rowcolor{gray!20} 
& 
& 
\texttt{/* tagged = tag \& \&(new\_A-$>$A[0]), }\\

\rowcolor{gray!20} 
&
&
{\tt A\_size = sizeof(int) * 10 */} \\ 

4 & 
\texttt{int *p;} & 
\texttt{int *p;} \\

5 & 
\texttt{p = A+idx;} & 
\texttt{p = tagged + idx;} \\

\rowcolor{gray!20} 
6& 
& 
\texttt{check\_inframe(tagged, p);} \\



\rowcolor{gray!20} 
7 & 
& 
\texttt{untagged\_p = }\\ 

\rowcolor{gray!20} 
&
&
\texttt{  check\_bounds(p, sizeof(int));}\\

8 & 
\texttt{*p = val;} & 
\texttt{*untagged\_p = val;}\\

\hline
\end{tabular}
}}
\hfill{}
\caption{\small{\pname{} inserts code, highlighted in gray, 
for creating a header-padded object, updating metadata 
and detecting memory corruption. Codes in line 2, 5,
and 8 in the first column are transformed to codes 
in the second column.}}
\label{tb:toycode}
\end{table}


\subsubsection{Memory allocation}

As described in Section~\ref{sec:transform_alloc},
a header is prepended to memory objects
(lines 1, 2 in Table~\ref{tb:toycode}). 
For spatial safety, this header must hold at least the raw object size,
but can hold additional information such as
a type id.  This could be used for additional
checks for sub-object bounds violations
or type confusion,
but we do not experiment with these in this work.

\begin{lstlisting}
struct HeaderTy {
  unsigned size;
  unsigned type_id; /* Suggested for other applications */
} 

struct __attribute__((packed)) newTy {
  HeaderTy hd; 
  Ty obj; /* Ty is an original object's type */
};
\end{lstlisting} 

An object's base address is obtained by adding 
{\tt sizeof(headerTy)} to the header address, 
once we get the header address from 
a tagged pointer.

Once a new object is allocated, an instrumented 
function ({\tt handle\_alloc}) updates 
metadata, moves the pointer to ({\tt new\_A->A}), 
and then tags it (line 3). The pointer to the removed 
original object is replaced with a tagged 
one ({\tt A} to {\tt tagged} in line 5).

\subsubsection{Pointer arithmetic}

Going out-of-bounds at pointer arithmetic is not 
memory corruption as long as they are not dereferenced. 
However, skipping checks at pointer arithmetic can 
lose track of pointers' \textit{intended referents}. 
Memory access to these pointer can be seen 
\textit{valid} in many object bounds-based approaches.  
To keep track of intended referents, 
object-tracking approaches may have to check bounds 
at pointer arithmetic~\cite{Jones97backwards-compatiblebounds}. 
However, performing bounds checks only at pointer arithmetic 
may therefore cause \textit{false positives}, where a pointer
going out-of-bounds by pointer arithmetic is not 
dereferenced as follows: 

\begin{lstlisting}
int *p;
int *a = malloc(n * sizeof(int));
for (p = a; p < &a[100]; p++) *p = 0;    
\end{lstlisting}

On the exit of {\tt for} loop, {\tt p} goes out-of-bounds 
yet is not dereferenced -- this is not an error in C standard. 
\cite{Akritidis:2009:BBC:1855768.1855772} handles this 
by marking the pointers during pointer arithmetic, 
and sending errors when dereferenced,
or padding an object by \textit{off-by-one byte}, 
causing most of the false 
positives~\cite{Jones97backwards-compatiblebounds}. 


Instead of padding, we include one \textit{imaginary} 
off-by-one byte (or multiple bytes) when deciding 
the wrapper frame (see Section~\ref{sec:frame_selection})
on memory allocation.
The fake padding then is within the wrapper frame, 
and pointers to this are still derived to the header,
even when they land another object by pointer arithmetic.
The biggest advantage of fake padding is that it is
allowed to be overlapped with neighboring objects.
The fake padding does not cause conflict 
of $N$ value with another object on the supplementary table 
possibly overlapping the bytes.

\pname{} tolerates pointers to the padding at 
pointer arithmetic, and reports errors on attempts 
to access them.    
\pname{} detects those pointers being dereferenced, 
since bounds checking at memory access retrieves 
the raw size of the object. 

Currently \pname{} adds fake padding only in the tail
of objects, but it could be also attached at the front
to track pointers going under lower bounds. 

Above utilizing fake padding,
to make a stronger guarantee for near-zero 
false negatives, we could perform 
\textit{in-frame checking} (currently disabled) 
at pointer arithmetic (line 6 in Table~\ref{tb:toycode}).
We can derive the header address of an intended referent, 
as long as the pointer stays inside its wrapper frame 
(slot for small-framed), in any circumstance. 
In Fig.~\ref{fig:in_frame_check}, consider a pointer 
({\tt p}), and its small-framed referent ({\tt a}).
Assuming {\tt p} going out-of-bounds to {\tt p'} by 
pointer arithmetic, {\tt p'} even 
violates its intended referent, but {\tt p'} 
is still within {\tt slot0}. Hence, {\tt p'} is
derived to {\tt a}'s header by zeroing 
lower $log_2 (slot\_size)$ (15) bits and 
adding {\tt offset}. This is
applied the same for big-framed objects.    

Hence, we could check only \textit{out-of-frame} ({\tt p"} 
in Fig.~\ref{fig:in_frame_check}) by performing 
simple bit-wise operations (no metadata retrieval)
checking if {\tt p} and {\tt p'} are in its wrapper 
frame (or slot for small-framed). 
 

\begin{lstlisting}
/* N: log2 wrapper_frame_size (or slot_size)*/ 
is_inframe = (p' ^ p) & (~0ULL << N); 
assert(is_inframe == 0); 
\end{lstlisting}




\tikzstyle{arrow}=[draw, -latex] 
  
\begin{figure}
\noindent\resizebox{\columnwidth}{!}{%
\begin{tikzpicture}[
    node distance=0pt,
    frame/.style={
        rounded corners=1mm,
        draw=black,
        font=\scriptsize\ttfamily, 
        minimum width=2cm,
        minimum height=0.3cm,
        inner sep=0pt
    },
    object/.style={
        rounded corners=1mm,
        draw=black,
        fill=white, 
        minimum width=0.6cm, 
        minimum height=0.2cm,
        font=\fontsize{6}{6}\ttfamily, 
        anchor=west,
    },
    leftflat/.style={
        append after command={%
            \pgfextra
            \fill[fill=#1] 
            (\tikzlastnode.north west) 
            [rounded corners] -| 
            (\tikzlastnode.east) 
            |- (\tikzlastnode.south) 
            [sharp corners] -| cycle;
        \draw[rounded corners] 
            (\tikzlastnode.north west) -| 
            (\tikzlastnode.east) |- 
            (\tikzlastnode.south west);
        \endpgfextra},
        minimum width=2cm,
        minimum height=0.3cm,
        inner sep=0pt,
        fill=bblue
    },
    rightflat/.style={
        append after command={%
            \pgfextra
            \fill[fill=#1] 
            (\tikzlastnode.south east) 
            [rounded corners] -| 
            (\tikzlastnode.west) |- 
            (\tikzlastnode.north) 
            [sharp corners] -| cycle;
        \draw[rounded corners] 
            (\tikzlastnode.south east) -| 
            (\tikzlastnode.west) |- 
            (\tikzlastnode.north east);
            \endpgfextra},
        minimum width=2cm,
        minimum height=0.3cm,
        font=\scriptsize\ttfamily, 
        inner sep=0pt
    },
    entry/.style={
        node distance=0pt,outer sep=0pt,
        draw=black,
        text width=#1,
        font={\sffamily\bfseries},
        align=center,
        minimum width=0.1cm,
        minimum height=0.3cm
    },
]
    \def\txtdist{0.35cm}
    \node[frame, 
        label=left:{\scriptsize\ttfamily $2^{14}$}] 
        at (0,1.5) (1) {};
    \node[frame,right=of 1] (2) 
        {};
    \node[frame,right=of 2] (3) 
        {};
    \widernode[frame, fill=pastelteal, 
        label=left:{\scriptsize\ttfamily $2^{15}$}]
        {1}{2}{\scriptsize\ttfamily slot0}{1_2};
    \node[rightflat=white,below=of 3] (4) 
        {\scriptsize\ttfamily slot1};
    \widernode[rightflat=white, 
        label=left:{\scriptsize\ttfamily $2^{16}$}]
        {1_2}{4}{\scriptsize\ttfamily 16-frame0}{1_4};

    \node (obj_a) [object, inner ysep=2.5pt] at (-0.15, 2.00) 
        {\textbf{a}};
    \node (obj_b) [object, inner ysep=2pt] at (0.8, 2.00) 
        {\textbf{b}};
  
   \node (ptr1) [inner ysep=-4pt]at (0, 2.5) {\scriptsize\ttfamily p};  
   \node (ptr2)  [inner ysep=-4pt] at (1.3, 2.5) {\scriptsize\ttfamily p'};  
   \node (ptr3)  [inner ysep=-4pt] at (3.4, 2.5) {\scriptsize\ttfamily p"};  
   
    \coordinate (ptr_1_start) at (0, 2.4);
    \coordinate (ptr_1_end) at (0, 2.15);
    \path [arrow] (ptr_1_start) -- (ptr_1_end);
    
    \coordinate (ptr_2_start) at (1.3, 2.4);
    \coordinate (ptr_2_end) at (1.3,2.15);
    \path [arrow] (ptr_2_start) -- (ptr_2_end);

    \coordinate (ptr_3_start) at (3.4, 2.4);
    \coordinate (ptr_3_end) at (3.4, 1.65);
    \path [arrow] (ptr_3_start) -- (ptr_3_end);
    
    \draw[|<->|] ([yshift=0.0cm] obj_a.west) -- node[above] 
        {\scriptsize\ttfamily offset} ([yshift=0.5cm]1.west);
    
    \end{tikzpicture}
}
\caption{
\small{By pointer arithmetic, 
a pointer {\tt p} goes out-of-bounds ({\tt p'}), and 
also violates its intended referent ({\tt a} to {\tt b}).  
\pname{} still can keep track of its referent, since {\tt p'}
is \textit{in-frame}. {\tt p"} is \textit{out-of-frame},
which we catch at pointer arithmetic.}}
\label{fig:in_frame_check}
\end{figure}
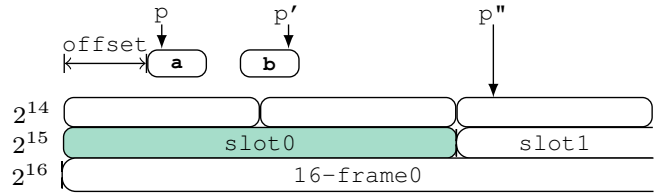


\pname{}'s only false positives are out-of-frame
pointers getting back in-frame without being dereferenced,
which is very rare. Those uses will be usually optimised 
away by compiler above optimisation level -O1, and 
normally the distance between an 
object's and its wrapper frame's bounds is large. 

\subsubsection{Memory access}

As mentioned in Section~\ref{sec:implementation_mem_access}, 
we instrument memory access
by replacing pointer operands with a return 
of our hook, so that the pointers are verified 
and tag-stripped, before being dereferenced 
(line 7,8 in Table~\ref{tb:toycode}).

{\tt check\_bounds} first reads a tagged pointer's flag 
telling if the object is small or big-framed. As we
described in Section~\ref{sec:small_frame} and~\ref{sec:big_frame}, 
we derive the header address from either an offset or
an entry, and then get an object's size from 
the header and its base address as follows:

\begin{lstlisting}
  obj_base = header_addr + sizeof(HeaderTy);
  obj_size = (HeaderTy *)header_addr->size;
\end{lstlisting}

We then check 
both under/overflows ((1) and (2) below, respectively).
Detection of underflows is essential for \pname{} to 
prevent overwrites to the header.

\begin{lstlisting}
assert(untagged_p >= obj_base);    /* (1) */
assert(untagged_p + sizeof(T) - 1
       <= upperbound));            /* (2) */
/* Where T is the type to be accessed */
\end{lstlisting}

The assertion (2) aims at catching overflows and memory 
corruption caused by access after unsafe typecast
such as the following example:

\begin{lstlisting}
char *p = malloc(10);
int *q = p + 8; 
*q = 10; // memory corruption 
\end{lstlisting}


In a similar fashion, we instrument {\tt memcpy, 
memmove, memset}, and string functions
({\tt strcpy, strncmp, strncpy, memcmp, memchr} and 
{\tt strncat}). Handling individual function 
depends on how each function works. 
For instance, {\tt strcpy} copies a string {\tt src} 
up to {\tt null}-terminated byte, and {\tt src}'s length may 
not be equal to the array size holding it. 
As long as the destination array is big enough to hold 
{\tt src}, it is safe, even if the source array is 
bigger than the destination array. 
Hence, we check if the destination size is not smaller 
than {\tt strlen(src)}, returning the length up to the 
{\tt null} byte as follows:

\begin{lstlisting}
char *__wrap_strcpy(char *dst, char *src)
{ /* Strip off a pointer's tag to pass it 
     to an external lib function (strlen) */
  char *untagsrc = (char *)untag_ptr(src); 
  size_t srclen = strlen(untagsrc);  
  
  /* Check if dest array size is not smaller
  than the lengh of src string */
  __real_strcpy((char *)check_bounds(dst, srclen), untagsrc);
  return dst; 
}
\end{lstlisting}

On the other hand, {\tt strncpy} copies a string up 
to user-specified n bytes, so we check both sizes of
destination and source arrays are bigger than n. Metadata
for both arrays are retrieved for bounds checking unlike
handling {\tt strcpy}.



\subsection{Temporal Memory Safety}\label{sec:dangling}

Although our primary focus in this paper is spatial safety,
\pname{} can also detect some forms of
temporal memory errors~\cite{Akritidis:2010:CMA:1929820.1929836,
1633516, cets, Simpson:2013:MES:2422144.2422147}
that we now discuss briefly.

Each big-framed object is mapped to an entry in a division array
in the supplementary table, and the entry is mapped to at 
most one big-framed object for each $N$. We make sure an 
entry is set to {\tt zero} whenever a corresponding object is
released. This way, we can detect
an attempt to {\tt free} an already deallocated object 
(i.e.~a \textit{double free}), by checking if the entry is zero.
Access to a deallocated object (i.e.~\textit{use-after-free})
is detected in the same way during metadata retrieval for a big-framed
object. Note that this cannot detect invalid \textit{temporal} 
intended referents, i.e., an object is released, 
a new object mapped to the same entry is allocated, 
and then a pointer attempts to access the first object.

Detection of dangling pointers for small-framed objects 
is out of scope of current implementation.


\subsection{Type Cast Checking}

The majority of type casts in C/C++ programs are either 
\textit{upcasts} (conversion from a descendant type to 
its ancestor type) or \textit{downcasts} (in the opposite 
direction). Upcasts are considered safe, and this can be 
verified at compile time, since if a source type of upcasts
is a descendant type, then the type of the allocated object
at runtime is also a descendant type. 

In contrast, the target type of downcasts may mismatch 
the run-time type (RTT). If an allocated object's type is a
descendant type of the target type at downcast,
access to an object after downcasts may cause 
boundary overruns including internal overflows. 
This is a vulnerability commonly known as 
\textit{type confusion}~\cite{typesan,hextype, 
Kell:2015:TDO:2814228.2814238, Kell:2016:DDT:2983990.2983998}.
The RTT is usually unknown statically due to 
inter-procedural data flows, so downcasts require 
run-time checking to prevent this type confusion.  

RTT verification is more challenging than 
upcast checking at compiler-time, since it requires 
\textit{pointer-to-type} mapping. We need to track individual 
objects (or pointers) and store per-object (pointer) type information 
in the database. In addition, RTT checking requires 
mappings of unique offsets to fields corresponding to types 
of sub-objects. 
\pname{} could be the basis of metadata storage (mapping
a pointer to per-object type) 
with supplementary type descriptors. 
\pname{}'s header can hold corresponding 
per-object type layout information (i.e.~a list of types at 
each offset in the object type) or its type ID 
for the object, and all type layout information 
and type-compatibility relations
can be stored in the type descriptors (implementations can vary).
\pname{}'s current implementation as an LTO pass makes it 
easier to collect all used types of the whole program.  


Downcasts may be critical for approaches using embedded
metadata (Section~\ref{sec:embedded_metadata}), 
since memory writes after unsafe type casts on program's 
user data can pollute metadata in a neighboring object's
header. Prevention of metadata corruption is easier with \pname{}
than with fat pointers.
We can detect memory overwrites to
another object's header caused by downcasts 
by simply keeping track of structure-typed objects and
using our bounds checking.   
Unlike fat pointers, we do not need to check internal
overflows by unsafe downcasts to protect metadata,
since metadata is placed outside an object.

\begin{center}
\begin{table*}[ht]
{\small
\hfill{}
\begin{tabular}{|l|c|c|c|c|c|c|c|c|}       
  \hline
                   & Memory    & Runtime  & Dynamic      & IPC  & Load     & D-cache  & Branch  & B-cache \\
                   & footprint & (cycles) & instructions &      & density  & MPKI     & density & MPKI  \\
  \hline
\textbf{Baseline}  &  1.00      & 1.00   & 1.00          & 1.70 & 0.28     & 24.85     & 0.19    & 2.85  \\
\textbf{Store-only}&  1.22      & 1.70   & 2.24          & 2.17 & 0.20     & 12.27     & 0.15    & 1.34  \\
\textbf{Full check}&  1.23      & 3.23   & 5.25          & 2.54 & 0.14     & 5.28     & 0.17    &  0.86 \\
\hline
\end{tabular}
\hfill{}
}
\caption{\small{Summary averages over all benchmarks (first three columns normalised)}}
\label{tb:summary_statisics}
\end{table*}
\end{center}


\section {Optimisations}\label{sec:optimisation}

We applied both our customised and LLVM built-in
optimisations. This section describes
our own optimisations. Suggestion of further  
optimisations is provided later in Section~\ref{sec:more_optimisation}. 

\paragraph{Implementation Considerations}

As described in Section~\ref{global_object_transformation},
all occurrences of an original pointer to a global 
object are replaced with a tagged pointer created using a 
constant expression (LLVM ConstExpr). Unfortunately, we experienced runtime 
hotspots due to the propagation of the tagging expression
to every ConstExpr using the original pointer. To work around this issue 
we created a helper global variable for each global object, assigned the constant 
tagged pointer to it during program initialisation,
and then replaced the occurrences of an original pointer 
with the corresponding helper variable. This way, runtime overheads are   
reduced. For instance, benchmark {\tt anagram}'s 
overhead decreased from 14 seconds to 1.7 seconds.  

\paragraph{Non-array Objects}

We do not track non-array objects that are not involved 
with pointer arithmetic, e.g., int-typed objects. 
It is redundant to perform bounds checking or un-tagging
for pointers to them. We filter out simple cases, 
easily recognised, from being checked.
In the general case, it is not trivial to determine 
if a pointer is untagged at compile time, 
since back-tracing the assignment for 
the pointer requires whole-program static analysis.

\paragraph{Safe Pointer Arithmetic}
Instead of full bounds checks, we only strip off tags
for pointers involved in pointer arithmetic and statically proven
in-bound. 
For pointers where the bounds can be determined statically, as in the following example,
we insert only runtime bounds checks and avoid the metadata retrieval:

\begin{lstlisting}
int a[10];
... *(a + n) ...
\end{lstlisting}



\paragraph{Hoist Run-time Checks Outside Loops}
\textit{Loop-invariant} expressions can be hoisted 
out of loops, 
thus improving run-time performance by executing the 
expression only once rather than at each iteration.
We modified SAFECode's~\cite{Dhurjati:2006:SEA:1133981.1133999, conf/isse/Simpson14}
loop optimisation passes.
We apply hoisting checks to monotonic loops, and pull
loop invariants that do not change throughout the loop, 
and scalars to the pre-header of each monotonic loop.
While iterating our run-time checks inside each loop 
including inner loops, we determine if the pointer 
is hoistable. If hoistable,
we place a scalar evolution expression along with its 
run-time checks outside the loop, and delete the old
checks inside loop.

\paragraph{Inlining Function Calls in the Loop}
Inlining functions can improve performance, 
however it can bring more performance degradation
due to the bigger size of the code (runtime checks
are called basically at every memory access). 
Currently, we only inline bounds checks that are 
inside loops to minimise code size.


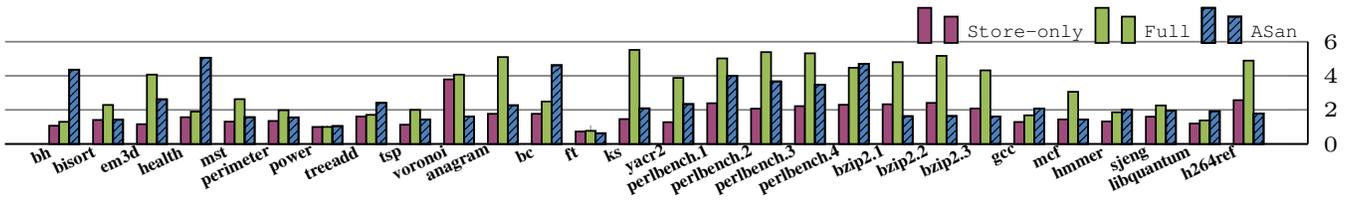
\begin{figure*}[ht]
\centering
\resizebox{\textwidth}{!}{%
    \begin{tikzpicture}
    \begin{axis}[
        xlabel={},
        ylabel={},
        axis x line*=bottom,
        axis y line*=right,
        enlarge x limits=0.05,
        set layers,
        ybar=0.0pt,
        bar width=0.08cm,
        width=12cm, height=2.4cm, 
        ymin=0, ymax=6,
        ytick={0,2,4,6},
        restrict y to domain*=0:6,
        symbolic x coords={bh, bisort, em3d, health, mst, 
            perimeter, power, treeadd, tsp, voronoi,
            anagram, bc, ft, ks, yacr2,
            perlbench.1, perlbench.2, perlbench.3, 
            perlbench.4, bzip2.1, bzip2.2, bzip2.3, gcc, 
            mcf, hmmer, sjeng, libquantum,h264ref},
        xtick={bh, bisort, em3d, health, mst, 
            perimeter, power, treeadd, tsp, voronoi,
            anagram, bc, ft, ks, yacr2,
            perlbench.1, perlbench.2, perlbench.3, 
            perlbench.4, bzip2.1, bzip2.2, bzip2.3, 
            gcc, mcf, hmmer, sjeng, libquantum,h264ref},
        visualization depends on=rawy\as\rawy,
        clip=false,
        legend cell align=left,
        x label style={font=\tiny},
        y label style={font=\tiny},
        xticklabel style={font=\fontsize{4}{4}\bfseries,rotate=25,anchor=east},
        yticklabel style={font=\fontsize{5}{5}\bfseries},
        ylabel near ticks,
        ymajorgrids, major grid style={draw=gray},
        legend style={
            draw=none,
            fill=none,
          at={(0.85, 0.8)},
          anchor=south, legend columns=-1, 
          font=\fontsize{4.5}{4.5}\ttfamily
        },
        xtick align=inside,
      ]
          \addplot+[black,fill=ppurple, font=\tiny] 
          coordinates {
                  (bh,	        1.07	)
                  (bisort,	    1.41	)
                  (em3d,	    1.16	)
                  (health,	    1.57	)
                  (mst,	        1.31	)
                  (perimeter,	1.35	)
                  (power,	    1.00	)
                  (treeadd,	    1.61	)
                  (tsp,	        1.13	)
                  (voronoi,	    3.79	)
                  (anagram,	    1.77	)
                  (bc,	        1.77	)
                  (ft,	        0.73	)
                  (ks,	        1.46	)
                  (yacr2,	    1.28	)
                  (perlbench.1,	2.39	)
                  (perlbench.2,	2.07	)
                  (perlbench.3,	2.22	)
                  (perlbench.4,	2.30	)
                  (bzip2.1,	    2.33	)
                  (bzip2.2,	    2.41	)
                  (bzip2.3,	    2.08	)
                  (gcc,	        1.29	)
                  (mcf,	        1.44	)
                  (hmmer,	    1.32	)
                  (sjeng,	    1.60	)
                  (libquantum,	1.20	)
                  (h264ref,	    2.57	)
        };
        \addplot[fill=ggreen, draw=black, font=\tiny] coordinates {
            (bh,	1.30	)
                (bisort,	2.29	)
                (em3d,	4.07	)
                (health,	1.90	)
                (mst,	2.63	)
                (perimeter,	1.98	)
                (power,	1.00	)
                (treeadd,	1.71	)
                (tsp,	2.00	)
                (voronoi,	4.07	)
                (anagram,	5.10	)
                (bc,	2.49	)
                (ft,	0.77	)
                (ks,	5.52	)
                (yacr2,	3.88	)
                (perlbench.1,	5.02	)
                (perlbench.2,	5.39	)
                (perlbench.3,	5.32	)
                (perlbench.4,	4.47	)
                (bzip2.1,	4.80	)
                (bzip2.2,	5.17	)
                (bzip2.3,	4.32	)
                (gcc,	1.68	)
                (mcf,	3.06	)
                (hmmer,	1.86	)
                (sjeng,	2.26	)
                (libquantum,	1.38	)
                (h264ref,	4.89	)
        };
        \addplot[fill=bblue, draw=black, 
                postaction={pattern=north east lines}, font=\tiny] coordinates {
                    (bh,	4.36	)
                        (bisort,	1.42	)
                        (em3d,	2.62	)
                        (health,	5.06	)
                        (mst,	1.57	)
                        (perimeter,	1.56	)
                        (power,	1.05	)
                        (treeadd,	2.41	)
                        (tsp,	1.43	)
                        (voronoi,	1.61	)
                        (anagram,	2.27	)
                        (bc,	4.62	)
                        (ft,	0.63	)
                        (ks,	2.09	)
                        (yacr2,	2.34	)
                        (perlbench.1,	3.99	)
                        (perlbench.2,	3.67	)
                        (perlbench.3,	3.48	)
                        (perlbench.4,	4.70	)
                        (bzip2.1,	1.63	)
                        (bzip2.2,	1.64	)
                        (bzip2.3,	1.61	)
                        (gcc,	2.08	)
                        (mcf,	1.44	)
                        (hmmer,	2.02	)
                        (sjeng,	1.95	)
                        (libquantum,	1.92	)
                        (h264ref,	1.79	)
                        };
      \legend{Store-only, Full, ASan}
      \end{axis}
    \end{tikzpicture}
  }
  \vspace*{-6mm}
  \caption{\small{Normalised runtime overheads}}
  \label{figure:overall_overhead}
\end{figure*}


\begin{figure*}[ht]
\centering
\resizebox{\textwidth}{!}{%
  \begin{tikzpicture}
    \begin{axis}[
        xlabel={},
        ylabel={},
        axis x line*=bottom,
        axis y line*=right,
        enlarge x limits=0.05,
        set layers,
        ybar=0.0pt,
        bar width=0.08cm,
        width=12cm, height=2.4cm, 
        ymin=0, ymax=10, 
        ytick={1, 2, 5, 10},
        restrict y to domain*=0:10,
        symbolic x coords={bh, bisort, em3d, health, mst, 
            perimeter, power, treeadd, tsp, voronoi,
            anagram, bc, ft, ks, yacr2,
            perlbench.1, perlbench.2, perlbench.3, perlbench.4, bzip2.1, 
            bzip2.2, bzip2.3, gcc, mcf, hmmer, sjeng,
            libquantum,h264ref},
        xtick={bh, bisort, em3d, health, mst, 
            perimeter, power, treeadd, tsp, voronoi,
            anagram, bc, ft, ks, yacr2,
            perlbench.1, perlbench.2, perlbench.3, perlbench.4, bzip2.1, 
            bzip2.2, bzip2.3, gcc, mcf, hmmer, sjeng,
            libquantum,h264ref},
        visualization depends on=rawy\as\rawy,
        after end axis/.code={ 
        },
        clip=false,
        legend cell align=left,
        x label style={},
        y label style={},
        xticklabel style={font=\fontsize{4}{4}\bfseries,rotate=25,anchor=east},
        yticklabel style={font=\fontsize{4}{4}\bfseries},
        ylabel near ticks,
        ymajorgrids, major grid style={draw=gray},
        legend style={
            at={(current bounding box.south-|current axis.south)},
            anchor=north, legend columns=-1, font=\tiny
        }, 
        xtick align=inside,
        ]
        \addplot[fill=ppurple, font=\tiny] coordinates {
                (bh,	1.10	)
                (bisort,	1.49	)
                (em3d,	1.01	)
                (health,	1.41	)
                (mst,	1.00	)
                (perimeter,	1.25	)
                (power,	1.19	)
                (treeadd,	1.51	)
                (tsp,	1.26	)
                (voronoi,	1.10	)
                (anagram,	1.05	)
                (bc,	1.04	)
                (ft,	1.31	)
                (ks,	1.11	)
                (yacr2,	2.14	)
                (perlbench.1,	1.18	)
                (perlbench.2,	1.38	)
                (perlbench.3,	1.21	)
                (perlbench.4,	1.36	)
                (bzip2.1,	1.21	)
                (bzip2.2,	1.29	)
                (bzip2.3,	1.03	)
                (gcc,	1.23	)
                (mcf,	1.02	)
                (hmmer,	1.08	)
                (sjeng,	1.04	)
                (libquantum,	1.15	)
                (h264ref,	1.10	)        };
    \addplot[fill=ggreen, draw=black, font=\tiny] coordinates {
        (bh,	1.10	)
            (bisort,	1.49	)
            (em3d,	1.01	)
            (health,	1.39	)
            (mst,	1.00	)
            (perimeter,	1.25	)
            (power,	1.20	)
            (treeadd,	1.51	)
            (tsp,	1.26	)
            (voronoi,	1.09	)
            (anagram,	1.06	)
            (bc,	1.05	)
            (ft,	1.31	)
            (ks,	1.12	)
            (yacr2,	2.16	)
            (perlbench.1,	1.12	)
            (perlbench.2,	1.84	)
            (perlbench.3,	1.15	)
            (perlbench.4,	1.24	)
            (bzip2.1,	1.07	)
            (bzip2.2,	1.14	)
            (bzip2.3,	1.01	)
            (gcc,	1.46	)
            (mcf,	1.00	)
            (hmmer,	1.10	)
            (sjeng,	1.00	)
            (libquantum,	1.15	)
            (h264ref,	1.14	)
    };
         \addplot[fill=bblue, 
                 draw=black, postaction={pattern=north east lines}, font=\tiny] coordinates {
                        (bh,	    3.47	)
                         (bisort,	2.25	)
                         (em3d,	    1.67	)
                         (health,	2.21	)
                         (mst,	    1.66	)
                         (perimeter,1.26	)
                         (power,	8.54	)
                         (treeadd,	1.76	)
                         (tsp,	    1.83	)
                         (voronoi,	1.43	)
                         (anagram,	14.02	) %
                         (bc,	    20.25	) %
                         (ft,	    2.10	)
                         (ks,	    16.59	) %
                         (yacr2,	8.14	)
                         (perlbench.1,	15.11	) %
                         (perlbench.2,	29.08	) %
                         (perlbench.3,	12.46	) %
                         (perlbench.4,	3.79	)
                         (bzip2.1,	1.74	)
                         (bzip2.2,	2.27	)
                         (bzip2.3,	1.13	)
                         (gcc,	    17.47	) %
                         (mcf,	    1.85	)
                         (hmmer,	48.66	) %
                         (sjeng,	1.08	)
                         (libquantum,	17.13	) %
                         (h264ref,	8.67	)
        };
        \pgfplotsset{
          after end axis/.code={
            \node[above] at (axis cs:anagram,8.8){\tiny{14.0}};
            \node[above] at (axis cs:bc,8.8){\tiny{20.3}};
            \node[above] at (axis cs:ks,8.8){\tiny{16.6}};
            \node[above] at (axis cs:perlbench.1,8.8){\tiny{15.1}};
            \node[above] at (axis cs:perlbench.2,8.8){\tiny{29.1}};
            \node[above] at (axis cs:perlbench.3,8.8){\tiny{12.5}};
            \node[above] at (axis cs:gcc,8.8){\tiny{17.5}};
            \node[above] at (axis cs:hmmer,8.8){\tiny{48.7}};
            \node[above] at (axis cs:libquantum,8.8){\tiny{17.1}};
          }
        }
      \end{axis}
    \end{tikzpicture}
}
  \vspace*{-6mm}
  \caption{\small{Normalised maximum resident set size}}
  \label{figure:memory_footprint}
\end{figure*}
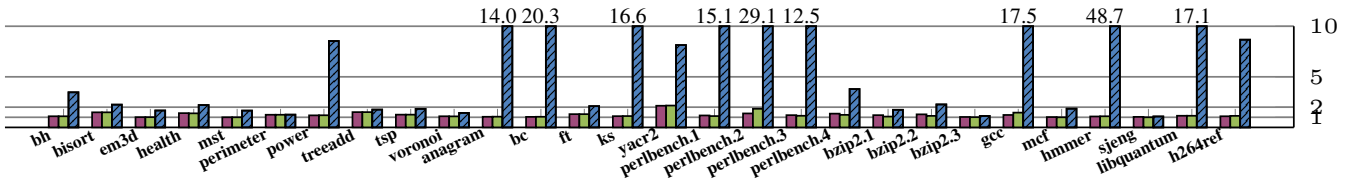

\section{Evaluation}\label{sec:evaluation}

We measured the performance of \pname{} on C benchmarks 
from the full set of Olden~\cite{Carlisle:1995:SCC:209936.209941}, 
Ptrdist~\cite{ptrdist}, 
and a subset of SPEC CPU 2006~\cite{Henning:2006:SCB:1186736.1186737}.
For each benchmark we measured four binary versions: 
un-instrumented, only store-checked and full 
(both load and store checking enabled) on \pname{},
and ASan. The same set of compiler optimisations 
in the same order are applied to four versions.
In addition, we disabled ASan's memory leak detection 
at run-time and halting on error in order to 
force ASan to continue after error detection.
This is to measure overheads in the same setting as \pname{}.
Binaries were compiled with the regular {\tt LLVM-clang} 
version 4.0 using optimisation level {\tt -O2}. 
Measurements were taken on an Intel\textsuperscript{\textregistered}
Xeon\textsuperscript{\textregistered}
E5-2687W v3 CPU with 132 GB of RAM.
Results were gathered using {\tt perf}.
Table~\ref{tb:summary_statisics} summarises 
the average of metrics of the baseline and the two
instrumented tests.

In this text, cache and branch misses 
refer to L1 D-cache misses and branch prediction misses 
both per 1000 instructions (MPKI), respectively. 

\subsection{Memory Overhead} 
Our metadata header was a generous 16 bytes per object.
The big-frame array had 48 elements for each 
16-frame (division) in use where the element size 
was 8 bytes to hold full address of the header. 
The header size and the number of 
elements of each division array can be reduced, but 
this is our early implementation.
Currently all the header objects were aligned by
16 for compatibility with the {\tt llvm.memset} 
intrinsic function that sometimes assumes this alignment.
Despite inflation of space using larger than needed 
headers and division array entries and some changes 
of alignment, we see \pname{}'s space overheads 
are very low at 1.22 and 1.23 as shown in Fig.~\ref{figure:memory_footprint}.
These measurements reflect code inflation for instrumenting 
both loads and stores.

The memory overheads of \pname{} are low and stable
compared to other approaches~\cite{softbound, address-sanitizer}.
ASan's average normalised overheads are 8.84 
(increase by 784\%) for the same working set
in our experiments, and the highest and lowest overheads  
are 4766\% for the test {\tt hmmer} and 8\% for 
{\tt sjeng}, respectively. The average memory overhead 
of \pname{} is 22\% $\sim$ 23\% for both store-only and full 
checking, and only two tests, {\tt perlbench.2} (84\%) and
{\tt yacr2} (116\%) recorded comparably higher growth 
than other tests.
The two tests produce many small-sized objects, 
for example, {\tt perlbench} allocates many 1-byte-sized 
heap objects. Currently \pname{}
instruments every heap object, so attaching a 
16-byte-sized header to all the 1-byte-sized objects 
made the increase higher. 
\pname{}'s overheads for those benchmarks are still 
much lower than ASan's: 2808\% for {\tt perlbench.2} 
and 714\% for {\tt yarc2}. 


%

\subsection{Slowdown} 

Fig.~\ref{figure:overall_overhead} reports the slowdown 
per benchmark (relative number of additional cycles). 
The average is 70\% for store-only and 223\% for full 
checking.

\begin{figure*}[ht]
\centering
\resizebox{\textwidth}{!}{%
    \begin{tikzpicture}
    \begin{axis}[
        xlabel={},
        ylabel={},
        axis x line*=bottom,
        axis y line*=right,
        set layers,
        enlarge x limits=0.05,
        ybar=0.1pt,
        bar width=0.08cm,
        width=12cm, height=2.4cm, 
        ymin=0, ymax=10,
        ytick={2,4,6,8,10},
        restrict y to domain*=0:10,
        symbolic x coords={bh, bisort, em3d, health, mst, 
            perimeter, power, treeadd, tsp, voronoi,
            anagram, bc, ft, ks, yacr2,
            perlbench.1, perlbench.2, perlbench.3, perlbench.4, bzip2.1, 
            bzip2.2, bzip2.3, gcc, mcf, hmmer, sjeng,
            libquantum,h264ref},
        xtick={bh, bisort, em3d, health, mst, 
            perimeter, power, treeadd, tsp, voronoi,
            anagram, bc, ft, ks, yacr2,
            perlbench.1, perlbench.2, perlbench.3, perlbench.4, bzip2.1, 
            bzip2.2, bzip2.3, gcc, mcf, hmmer, sjeng,
            libquantum,h264ref},
        visualization depends on=rawy\as\rawy,
        clip=false,
        legend cell align=left,
        x label style={font=\tiny},
        y label style={font=\tiny},
        xticklabel style={font=\fontsize{4}{4}\bfseries,rotate=25,anchor=east},
        yticklabel style={font=\fontsize{5}{5}\ttfamily},
        every node near coord/.append style={rotate=90, anchor=west},
        ylabel near ticks,
        ymajorgrids, major grid style={draw=gray},
        legend style={
            at={(current bounding box.south-|current axis.south)},
            anchor=north, legend columns=-1, font=\tiny
        },
        xtick align=inside,
        ]
        \addplot[black, ybar, fill=ppurple, 
         font=\tiny] coordinates {
                (bh,	    1.40	)
                 (bisort,	2.56	)
                 (em3d,	    1.59	)
                 (health,	2.03	)
                 (mst,	    2.26	)
                 (perimeter,1.50	)
                 (power,	0.82	)
                 (treeadd,	1.66	)
                 (tsp,	    1.44	)
                 (voronoi,	2.84	)
                 (anagram,	2.36	)
                 (bc,	    1.90	)
                 (ft,	    1.70	)
                 (ks,	    1.70	)
                 (yacr2,	1.54	)
                 (perlbench.1,	2.60	)
                 (perlbench.2,	2.39	)
                 (perlbench.3,	2.24	)
                 (perlbench.4,	2.66	)
                 (bzip2.1,	3.79	)
                 (bzip2.2,	4.00	)
                 (bzip2.3,	3.52	)
                 (gcc,	    1.49	)
                 (mcf,	    3.91	)
                 (hmmer,	1.93	)
                 (sjeng,	2.05	)
                 (libquantum,1.43	)
                 (h264ref,	3.40	)
        };
         \addplot[fill=ggreen, draw=black, font=\tiny] coordinates {
                 (bh,	    2.94	)
                 (bisort,	5.57	)
                 (em3d,	    5.93	)
                 (health,	3.27	)
                 (mst,	    6.92	)
                 (perimeter,2.76	)
                 (power,	0.87	)
                 (treeadd,	1.95	)
                 (tsp,	    5.77	)
                 (voronoi,	3.72	)
                 (anagram,	7.74	)
                 (bc,	    2.81	)
                 (ft,	    6.99	)
                 (ks,	    9.12	)
                 (yacr2,	5.18	)
                 (perlbench.1,	5.09	)
                 (perlbench.2,	5.58	)
                 (perlbench.3,	5.54	)
                 (perlbench.4,	5.29	)
                 (bzip2.1,	8.56	)
                 (bzip2.2,	8.84	)
                 (bzip2.3,	8.56	)
                 (gcc,	    1.85	)
                 (mcf,	    11.98	) %
                 (hmmer,	2.31	)
                 (sjeng,	3.29	)
                 (libquantum,1.58	)
                 (h264ref,	6.87	)         
         };
         \addplot[fill=bblue, draw=black, postaction={pattern=north east lines}, font=\tiny] coordinates {
                 (bh,	    14.36	) %
                 (bisort,	2.25	)
                 (em3d,	    2.70	)
                 (health,	3.74	)
                 (mst,	    2.28	)
                 (perimeter,2.38	)
                 (power,	1.45	)
                 (treeadd,	2.97	)
                 (tsp,	    2.44	)
                 (voronoi,	1.88	)
                 (anagram,	3.06	)
                 (bc,	    4.24	)
                 (ft,	    2.98	)
                 (ks,	    2.81	)
                 (yacr2,	2.91	)
                 (perlbench.1,	3.66	)
                 (perlbench.2,	3.33	)
                 (perlbench.3,	2.85	)
                 (perlbench.4,	4.44	)
                 (bzip2.1,	2.56	)
                 (bzip2.2,	2.58	)
                 (bzip2.3,	2.61	)
                 (gcc,	    2.28	)
                 (mcf,	    3.16	)
                 (hmmer,	3.75	)
                 (sjeng,	2.85	)
                 (libquantum,2.14	)
                 (h264ref,	2.57	)
       }; 
        \pgfplotsset{
          after end axis/.code={
            \node[above] at (axis cs:mcf,8.8){\tiny{11.97}};
            \node[above] at (axis cs:bh,8.8){\tiny{14.42}};
          }
        }
      \end{axis}
    \end{tikzpicture}
}
  \vspace*{-6mm}
  \caption{\small{Normalised dynamic instruction count}}
  \label{figure:dynamic_instruction_count}
\end{figure*}
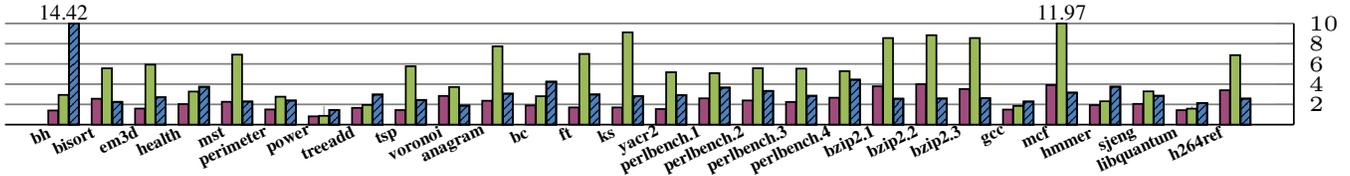

Our performance degradation is mainly due to increased dynamic
instructions. For full-checking, {\tt anagram} (410\%) 
and {\tt ks} (452\%) stand out for high overheads 
despite its smaller program size, mostly 
due to heavy recursion and excessive allocations causing 
big growth in executed instructions (674\% for {\tt anagram}, 
812\% for {\tt ks}) as shown in Fig.~\ref{figure:dynamic_instruction_count}, 
but decreases in cache misses 
are moderate (76\% for {\tt anagram}, 81\% for {\tt ks}) 
compared to average (decreased by 63\%). 
On contrast,
{\tt mcf} recorded the highest instruction overheads 
(1097\%), but cache misses (91\%) and branch misses 
(92\%) after instrumentation are dropped the biggest 
among all the tests, so run-time overhead did not grow 
in proportion to increased instruction count.   
{\tt perlbench} and {\tt bzip} sets' overheads are high in
both \pname{} and ASan, and ASan recorded better speed
in the sets ({\tt perlbench}: 405\% (\pname{}) and 299\% (ASan), 
and {\tt bzip}: 376\% (\pname{}) and 63\% (ASan)),
and especially in {\tt bzip} group. Both {\tt perlbench}
and {\tt bzip} produce many objects, and especially {\tt bzip}
recorded much higher growth in executed instructions
for metadata updates and checks than {\tt perlbench}
and others.

ASan (139\%) showed better performance than 
\pname{} (223\%) on the full-checking mode on average. 
\pname{} was faster than ASan for 9 tests among 28.

Performance was 
impacted far less than would naively be expected from 
the additional dynamic instruction count (metric 
columns 2 and 3 in Table~\ref{tb:summary_statisics}). 
The rise in IPC (column 4) 
is quite considerable on average, 
although the figure varies greatly by benchmark.
The original IPC ranged from 0.22 to 3.20 but after 
instrumentation there was half as much variation.





 


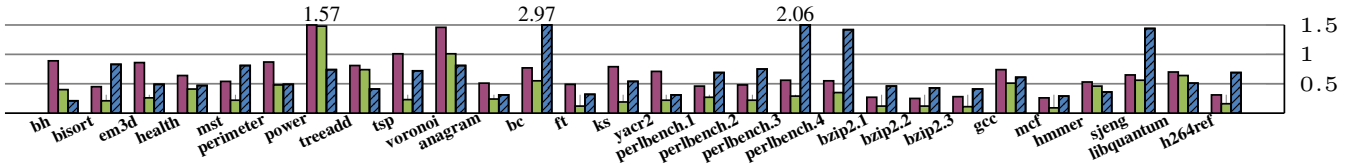
\begin{figure*}[ht]
\centering
\resizebox{\textwidth}{!}{%
    \begin{tikzpicture}
      \begin{axis}[
         xlabel={},
        ylabel={},
        axis x line*=bottom,
        axis y line*=right,
        enlarge x limits=0.05,
        set layers,
        ybar=0.0pt,
        bar width=0.08cm,
        width=12cm, height=2.3cm, 
        ymin=0, ymax=1.5,
        ytick={0.5,1,1.5},
        restrict y to domain*=0:1.5,
        symbolic x coords={bh, bisort, em3d, health, mst, 
            perimeter, power, treeadd, tsp, voronoi,
            anagram, bc, ft, ks, yacr2,
            perlbench.1, perlbench.2, perlbench.3, perlbench.4, bzip2.1, 
            bzip2.2, bzip2.3, gcc, mcf, hmmer, sjeng,
            libquantum,h264ref},
        xtick={bh, bisort, em3d, health, mst, 
            perimeter, power, treeadd, tsp, voronoi,
            anagram, bc, ft, ks, yacr2,
            perlbench.1, perlbench.2, perlbench.3, perlbench.4, bzip2.1, 
            bzip2.2, bzip2.3, gcc, mcf, hmmer, sjeng,
            libquantum,h264ref},
        visualization depends on=rawy\as\rawy,
        after end axis/.code={ 
            \draw [thick, white, decoration={snake, amplitude=1pt}, decorate] (rel axis cs:0,1.05) -- (rel axis cs:1,1.05);
        },
        clip=false,
        x label style={font=\tiny},
        y label style={font=\tiny},
        xticklabel style={font=\fontsize{4}{4}\bfseries,rotate=25,anchor=east},
        yticklabel style={font=\fontsize{5}{5}\ttfamily},
        ylabel near ticks,
        ymajorgrids, major grid style={draw=gray},
        legend style={
            draw=none,
            fill=none,
          at={(0.85, 0.8)},
          anchor=south, legend columns=-1, 
          font=\fontsize{4.5}{4.5}\ttfamily
        },
        xtick align=inside,
        ]
        \addplot[fill=ppurple, font=\tiny] coordinates {
                (bh,	    0.89	)
                (bisort,	0.45	)
                (em3d,	    0.86	)
                (health,	0.64	)
                (mst,	    0.54	)
                (perimeter,	0.87	)
                (power,	    1.57	)  %
                (treeadd,	0.81	)
                (tsp,	    1.01	)
                (voronoi,	1.46	)
                (anagram,	0.51	)
                (bc,	    0.77	)
                (ft,	    0.49	)
                (ks,	    0.79	)
                (yacr2,	    0.71	)
                (perlbench.1,	0.46	)
                (perlbench.2,	0.48	)
                (perlbench.3,	0.56	)
                (perlbench.4,	0.55	)
                (bzip2.1,	0.27	)
                (bzip2.2,	0.25	)
                (bzip2.3,	0.28	)
                (gcc,	    0.74	)
                (mcf,	    0.26	)
                (hmmer,	    0.53	)
                (sjeng,	    0.65	)
                (libquantum,0.70	)
                (h264ref,	0.31	)
        };
        \addplot[fill=ggreen, draw=black, font=\tiny] coordinates {
                (bh,	    0.40	)
                (bisort,	0.21	)
                (em3d,	    0.26	)
                (health,	0.41	)
                (mst,	    0.22	)
                (perimeter,	0.48	)
                (power,	    1.48	)
                (treeadd,	0.74	)
                (tsp,	    0.23	)
                (voronoi,	1.01	)
                (anagram,	0.24	)
                (bc,	    0.55	)
                (ft,	    0.12	)
                (ks,	    0.19	)
                (yacr2,	    0.22	)
                (perlbench.1,	0.27	)
                (perlbench.2,	0.22	)
                (perlbench.3,	0.29	)
                (perlbench.4,	0.35	)
                (bzip2.1,	0.12	)
                (bzip2.2,	0.12	)
                (bzip2.3,	0.11	)
                (gcc,	    0.51	)
                (mcf,	    0.09	)
                (hmmer,	    0.46	)
                (sjeng,	    0.56	)
                (libquantum,0.64	)
                (h264ref,	0.16	)
        };
        \addplot[fill=bblue, draw=black, postaction={pattern=north east lines}, font=\tiny] coordinates {
                (bh,	    0.21	)
                (bisort,	0.83	)
                (em3d,	    0.49	)
                (health,	0.47	)
                (mst,	    0.81	)
                (perimeter,	0.49	)
                (power,	    0.74	)
                (treeadd,	0.41	)
                (tsp,	    0.72	)
                (voronoi,	0.81	)
                (anagram,	0.31	)
                (bc,	    2.97	) %
                (ft,	    0.32	)
                (ks,	    0.54	)
                (yacr2,	    0.31	)
                (perlbench.1,	0.69	)
                (perlbench.2,	0.75	)
                (perlbench.3,	2.06	) %
                (perlbench.4,	1.42	)
                (bzip2.1,	0.46	)
                (bzip2.2,	0.43	)
                (bzip2.3,	0.41	)
                (gcc,	    0.61	)
                (mcf,	    0.29	)
                (hmmer,	    0.36	)
                (sjeng,	    1.44	)
                (libquantum,0.51	)
                (h264ref,	0.69	)        
        };
        \pgfplotsset{
          after end axis/.code={
            \node[above] at (axis cs:power,1.3) {\tiny{1.57}};
            \node[above] at (axis cs:bc,1.3)    {\tiny{2.97}};
            \node[above] at (axis cs:perlbench.3,1.3){\tiny{2.06}};
        }
        }
      \end{axis}
    \end{tikzpicture}
}
  \vspace*{-6mm}
  \caption{\small{Normalised L1 D-cache load misses per 1000 instructions (MPKI)}}
  \label{figure:l1_dcache_misses}
\end{figure*}



\subsection{Data Cache Misses} 
A primary goal of \pname{} is to allow flexible 
relationships between object and header locality 
so that additional cache misses from 
metadata access can be minimised.
We do not analyse L1 instruction cache miss 
rate since this generally has negligible
performance effect on modern processors, despite 
our slightly inflated code. To explain the measured 
increase in IPC we analyse L1 D-cache misses MPKI 
(cache misses) and branch prediction misses MPKI.
The baseline D-cache miss rate was 2.48\% 
(Table~\ref{tb:summary_statisics}) but this improves
with \pname{} enabled owing to repeated access to 
the same cache data.


In Fig.~\ref{figure:l1_dcache_misses}, we normalise 
cache misses to the uninstrumented figure.
The average normalised cache misses is 
0.66 and 0.38 for store-only and full-checking,
respectively.
The miss rate is reduced since the additional operations 
we add have high cache affinity 
which dilutes the underlying miss rate of the application.
Moreover, cache misses after instrumentation 
do not increase in proportion to increase in the 
number of dynamic instruction.

In the full-check mode, cache misses increased only for 
two test (48\% for {\tt power} and 1\% for {\tt voronoi}) 
on \pname{}, 
while ASan showed increase for four tests. ASan's normalised 
misses on average is 0.73, which is higher than \pname{}'s
0.38. ASan's highest overhead is 197\% for {\tt bc}, 
and two tests reached increase more than by 100\%.
On \pname{}, {\tt power}'s overhead by 48\% is mainly caused 
by the very low increase in instruction executed in 
producing MPKI. The rest of benchmarks' misses except two 
tests decreased after instrumentation, and normalised misses 
were below 0.5 for 21 tests among 28 working set 
for the full checking. As for ASan, 
only 13 tests' normalised misses were lower than 0.5.
In the store-only mode of \pname{}, the decrease is 
lower than that of the full-checking, but still the half of tests' 
normalised misses are below 0.5.  
The overall cache miss rate showed \pname{} 
is cache-efficient and stable. 

\subsection{Instructions Executed} 

\pname{} increases dynamic instruction count 
by 124\% for store-only, and 425\% for full checking.
Along with additional data access for metadata manipulation, 
this increase in instructions executed is 
the main contributor to runtime overhead. 
Our dynamic instruction penalty arises from setting up and
using tagged pointers.
Certain operations are easily verified at compile time.
If all checking could be performed statically there would
be no runtime overhead. But \pname{} inserts tags on all
pointers and use sites that are readily checkable at compile
time still suffer run-time overhead from pointer stripping
operations since all major architectures require the top
bits to be zero (or special pointer authentication code 
in ARM8) to avoid a segmentation fault. In addition, 
we sometimes have to dynamically strip the tag field
even for untagged pointers, i.e.~pointers to objects 
not tracked for bounds checking when it is uncertain
if the pointer is tagged or not.
The penalty of using tagged pointers is \textit{over-instrumentation}
-- unless individual memory access is proven safe statically,
we have to instrument (i.e.~tag-cleaning) memory access 
to avoid segmentation fault.
Another major source of overheads is arithmetic operations
to calculate metadata locations.
This makes cache misses overheads per 1000 instructions
(MPKI) for full-checking lower than for store-only.

The average overhead for ASan is 226\%, 
which is lower than \pname{}.
The average excluding the highest test 
(1336\% for {\tt bh}) is 184\%, 
while \pname{}'s average excluding the highest
(1098\% for {\tt mcf}) is 400\%.
The difference of slowdown between \pname{} and ASan 
on average (\pname{}: 213\%, ASan: 139\%) 
was not big  as the difference of instruction
executed due to \pname{}'s cache efficiency.
 

In summary, ASan is more efficient in 
instructions executed. ASan saves more dynamic instructions 
with using non-tagged pointers (no tag cleaning) 
and shadow space-only metadata storage that helps 
simpler derivation of metadata location, 
as trade-off of high locality and space usage. 
  
Future implementations can optimise the case where
conservative analysis reveals the tag never needs to be added.
More discussion on optimisation is described in Section~\ref{sec:discussion_optimisation}.



\subsection{Branch Misses} 
Additional conditional branches arise in \pname{} from
checking whether a big or small frame is used and in
the pointer validity checks themselves.
Many approaches using shadow space such as ASan
may be more relieved from these branches at metadata
retrieval, and have them at bounds checking using 
retrieved metadata.  

As shown in Table~\ref{tb:summary_statisics} col 7, 
the dynamic branch density decreases slightly under 
\pname{} instrumentation, but the branch mis-prediction
rate greatly decreases (col 8).
The averages of normalised branch misses for store-only 
and full-checking are 0.62 and 0.42, respectively.
This shows the additional branches added achieve highly 
accurate branch prediction and that branch predictors 
are not being overloaded.
Of the new branches added, the ones checking small/large 
frame size are completely statically predictable owing to 
the checking code instances being associated with
a given object.  And the ones checking pointer validity 
also predict perfectly since
no out-of-bounds errors are detected.



\begin{center}
\begin{table*}[ht]
{\small
\hfill{}
\resizebox{\textwidth}{!}{
\begin{tabular}{|>{\columncolor{gray!20}}r|r|r|r|r|>{\columncolor{gray!20}}r|r|r|r|r|>{\columncolor{gray!20}}r|r|r|r|r|}       
\hline
Benchmarks&     CL&     \#I &   DC&    SP & 
Benchmarks&     CL&     \#I &   DC&    SP & 
Benchmarks&     CL&     \#I &   DC&    SP \\
\hline
\hline
bh&	        3.4 &	4.9	&	0.5	&	3.2	&
bisort&	    0.6 &	0.4	&	3.9	&	1.5	&
em3d&	    0.6 &	0.5	&	1.9	&	1.7	\\

health&	    2.7	&	1.1	&	1.2	&	1.6	&
mst&	    0.6	&	0.3	&	3.7	&	1.7	&
perimeter&	0.8	&	0.9	&	1.0	&	1.0	\\

power&	    1.1	&	1.7	&	0.5	&	7.2	&
treeadd&	1.4	&	1.5	&	0.6	&	1.2	&
tsp&	    0.7	&	0.4	&	3.1	&	1.5	\\

voronoi&	0.4	&	0.5	&	0.8	&	1.3	&
anagram&	0.4	&	0.4	&	1.3	&	13.2	&
bc&	        1.9	&	1.5	&	5.4	&	19.4	\\

ft&	        0.8	&	0.4	&	2.6	&	1.6	&
ks&	        0.4	&	0.3	&	2.9	&	14.8	&
yacr2&	    0.6	&	0.6	&	1.4	&	3.8	\\
\hline
perlbench.1&0.8	&	0.7	&	2.6	&	13.6	&
perlbench.2&0.7	&	0.6	&	3.4	&	15.8	&
perlbench.3&0.7	&	0.5	&	7.1	&	10.8	\\

perlbench.4&1.1	&	0.8	&	4.1	&	3.1	&
bzip2.1&	0.3	&	0.3	&	3.8	&	1.6	&
bzip2.2&	0.3	&	0.3	&	3.7	&	2.0	\\

bzip        2.3 &	0.4	&	0.3	&	3.6	&	1.1	&
gcc&	    1.2	&	1.2	&	1.2	&	12.0	&
mcf&	    0.5	&	0.3	&	3.3	&	1.9	\\

hmmer&	    1.1	&	1.6	&	0.8	&	44.4	&
sjeng&	    0.9	&	0.9	&	2.6	&	1.1	&
libquantum&	1.4	&	1.4	&	0.8	&	14.8	\\

h264ref&	0.4	&	0.4	&	4.3	&	7.6	&

\cellcolor{pastelteal}\textbf{AVERAGE}&
\cellcolor{pastelteal}\textbf{0.74 }&
\cellcolor{pastelteal}\textbf{0.62 }&
\cellcolor{pastelteal}\textbf{1.93 }&
\cellcolor{pastelteal}\textbf{7.19 }&
&&&&	\\
\hline

\end{tabular}
}}
\hfill{}
\caption{\small{Address Sanitizer's normalised overheads 
divided by \pname{}'s (full-check): the columns are cycles 
(CL), instructions executed (\#I), L1 D-cache misses 
in MPKI (DC), and memory footprint (SP), respectively. 
The average is highlighted in light green in the last line.}}
\label{tb:comp_with_asan}
\end{table*}
\end{center}


\section {Discussion}\label{sec:discussion}

\subsection{Comparison with Other Approaches}


\subsubsection{Address Sanitizer} 




First of all, ASan is integrated into Clang 
front end, whereas \pname{} is implemented as 
a LLVM pass. ASan's placement has better 
chances for minimal instrumentation 
and maximal optimisation of redundant checks. 

Taking advantage of shadow space,
ASan's calculation of metadata retrieval and bounds checking 
is simpler than \pname{}, saving executed instructions. 
ASan aligns an object to 8 to avoid conflicts in entries, 
and maps every 8 byte to its entry holding the first $k$
bytes that are addressable. This 
simplifies operations of bounds checking, 
but makes it difficult to detect un-aligned memory 
access after unsafe type cast. 
In addition, ASan pads each object 32 bytes 
at minimum for redzone 
and extra 31 bytes for alignment, which burdens space.
On contrast, \pname{}'s fake padding 
and wrapper frames do not consume space,
allowing invading other objects' territory.
Just mind that adding excessive fake padding 
may enlarge an object's wrapper frame size, turning 
a small-framed object into a big-framed one,
that requires indirect memory access to metadata.  

Table~\ref{tb:comp_with_asan} summarises the comparison 
of overheads between ASan and \pname{} per benchmark. 
\pname{} showed higher efficiency 
in space and L1 D-cache hit overall. ASan's memory footprint 
is higher than \pname{} for the whole 28 benchmarks: 
increase by 23\% for \pname{}'s full-checking 
and 784\% for ASan on average.
ASan is based on shadow space only, while
\pname{} is mainly header-based, so the evaluation
showed that \pname{} is more cache-friendly:     
ASan's cache misses are higher than \pname{}
except for six benchmarks, and normalised cache misses
on average are 0.38 for \pname{} and 0.73 for ASan. 
This shows that \pname{}'s decrease in cache misses 
after instrumentation is much bigger.


On average, ASan showed better performance at run-time 
performance and executed instructions. 
Normalised overheads of cycle are 3.23 for \pname{} and 2.39
for ASan, and overheads of executed instructions are
5.25 (\pname{}) and 3.26 (ASan).
Among the 28 benchmarks, \pname{} was faster for 9 tests, and 
showed less increase in dynamic instruction counts
for 8 tests. As shown in Table~\ref{tb:comp_with_asan}, 
the benchmarks where ASan is faster
roughly match those with less instructions executed.
 
\subsubsection{SGXBounds}

Both SGXBounds and \pname{} are based on tagged pointers.
The main difference is that 
SGXBounds uses 32 bits for a tag among 64 bits, 
while \pname{} tags 
only upper spare 16 bits that are common. 
SGXBounds is 
not applicable to many 64-bit machine. 

SGXBounds's retrieving an upper bound first 
not the base (lower bound)
like \pname{}, may save some overheads 
if we perform overflow-only checking, which is more common.
However, using a footer makes systems slightly 
more vulnerable to metadata pollution at the 
same time without near complete memory safety.
For both over/under underflow checking, we 
do not consider our derivation of the base, 
not the upper bound, as a weakness. 
In addition, \pname{}'s frame encoding can be easily
integrated to SGXBounds' design. 

\subsubsection{MPX}\label{sec:comp_mpx}
In principle, \pname{} could 
utilize MPX for a faster instrumentation when used 
for spatial safety. We showed \pname{} is more 
cache-friendly, but it could be made even faster 
if a single instruction implemented the complete tag 
decode operation, splitting apart the tagged pointer 
into an untagged object pointer and separate header 
pointer in another register. This would be a fairly 
simple, register-to-register instruction, 
operating on general purpose registers. Since this
has not used the D-cache, an enhancement would be to 
compare the pointer against a bounds
limit at hardcoded offset loaded from the header, 
but the best design requires further study.

\subsubsection{Baggy Bounds and Low-Fat pointers}
Baggy bounds~\cite{Akritidis:2009:BBC:1855768.1855772}
re-aligns objects and adds padding to them to prevents 
conflicts entries in shadow space. 
One of benefits of \pname{} is allowance of 
wrapper frames being \textit{overlapped}, which removes 
both superfluous padding and metadata conflicts. 
In a similar way, our frame-based encoding 
is less intrusive than Low-fat pointers'~\cite{Kwon13low-fatpointers:},
that requires objects to be grouped and aligned by their size. 

\subsection{Additional Optimisations}\label{sec:more_optimisation}

\subsubsection{Runtime checks only at pointer arithmetic}
We aligned objects by 16 bytes at the moment due to 
llvm.memset intrinsic function. On this alignment,
we have spare 4 bits at the end of offset for small-framed.
(We already have spare bits for big-framed ones.)
Using the bits, we can mark out-of-bounds pointers at 
pointer arithmetic, and report errors when they are 
dereferenced. 
This way, we expect to remove duplicated runtime checks, 
since the pointer may be used for memory access 
multiple times.
 
 
\subsubsection{Compiler Optimisation}\label{sec:discussion_optimisation}

There are more optimisations we could use.
Duplicate runtime checks can be eliminated using 
\textit{dominator tree}. 
SoftBound \cite{softbound} reported that 
their simple dominator-based redundant check 
elimination improved performance
by 13\% 
and claimed more advanced elimination~\cite{Bodik:2000:AEA:349299.349342, Wurthinger:2007:ABC:1294325.1294343}
can reduce more overheads. 

Loop optimization showed minor impact on reducing overheads, 
even for some SPEC benchmarks whose number of hoisted 
run-time checks reached hundreds at static time.
Our naive optimization skipping untagging 
improved performance more than state-of-the-art 
loop hoist pass.  
\textit{Static points-to analysis}
\cite{Steensgaard:1996:PAA:237721.237727,
Tan:2017:EPP:3062341.3062360}, 
as long as it does not assume
the absence of memory errors, potentially enables 
many tags and bounds checks to be removed at compile time. 



\subsection{Hardware Implementation of \pname{}}

We believe \pname{}'s encoding is at its best 
when it is implemented as instruction set extensions.
As briefly mentioned in~\ref{sec:comp_mpx}, 
the increase in the number of executed instructions, 
the main contributor to slowdown of \pname{}, 
can be resolved with new instructions.
Tag-cleaning would be just one instruction. Moreover,  
decision of wrapper frame on memory allocation or 
calculation of metadata locations from tagged pointers
can be implemented as a single instruction, respectively.

In addition, hardware-based \pname{} can overcome MPX's overheads.
MPX suffers from space overheads and even crashes due to 
the limited space to store
per-pointer metadata in the bounds table, especially
for pointer-intensive programs. \pname{}'s per-object
metadata in the header and compact-size table 
can reduce the space. 
\pname{}'s tracking objects also removes metadata 
updates at pointer assignment,
that causes performance loss of MPX.   

\section{Conclusion}

We designed, implemented and evaluated \pname{}, 
a software capability model with object granularity.
\pname{} enables a variety of security
applications for detecting memory safety errors.
Compared to existing approaches,
our frame-based offset encoding
is more flexibile both in metadata association 
and memory management, while still offering
a fairly simple calculation to map from 
arbitrary pointers to metadata locations.
In addition, its intrinsic memory and
cache-efficiency make it potentially attractive for  
direct hardware support.



{\footnotesize \bibliographystyle{acm}
\bibliography{JinsBib}}

\appendix

\paragraph{Proof 1}\label{appendix1}

Given an object $o$ and its wrapper frame $f$,
let's assume there exists a smaller frame $x$ 
that has $o$ inside. Since $o$ resides in 
both $f$ and $x$, we can conclude that $x$ is a subframe 
of $f$. According to the assumption, the base address
of $o$ ($base_{o}$) is within the range of $x$, 
hence, we get $base_{x} \leq base_{o}$. 
Here, $f$ is $o$'s wrapper frame, so $base_{o}$ 
is placed in $f$'s lower subframe.  
$x$ is a subframe of $f$, hence $x$ must 
be $f$'s lower subframe. 
This is resolved to contradiction between the assumption 
($x$ has $o$ inside) and the definition of wrapper function
($o$'s upper bound in the upper subframe).    
Hence, we can conclude that
there is no smaller frame than $o$'s 
wrapper frame; this is actually the unique wrapper frame,
and it can be used a reference point. 

\paragraph {Proof 2}\label{appendix2}
We prove that for each $N$, there exists at most 
one $N$-object mapped to each entry of a division array,
and show $N$ identifies an object mapped to the
same division array. To prove this, 
we assume there exist two distinctive
objects, $x$ and $y$; both are $N$-objects 
($N \geq 16$) mapped to the same division array. 
Since $x$ and $y$ are $N$-objects, their wrapper 
frame ($f_x$ and $f_y$) is $2^N$-sized 
by definition. The division is the only one 
that $f_x$ and $f_y$ are mapped to as shown 
previously, so $f_x$ and $f_y$ have the same
base address as the division. In addition, both frames
have the same size, so they are identical. Both base
addresses of $x$ and $y$ ($b_x, b_y$) must be in the lower 
$(N-1)$-subframe of $f_x$ (or $f_y$), and end 
addresses must be in the other sub-frame. From this,
$b_x$ and $b_y$ must be smaller than $e_x$ and 
$e_y$. However, the objects are distinct,
so $b_x<e_x<b_y<e_y$ or vice versa must hold. 
The assumption leads to a contraction. We 
conclude that for each $N$, there is a unique 
$N$-object mapped to one division array.


\end{document}